\documentclass[11pt]{iopart}
\usepackage[english]{babel}
\pdfoutput=1
\usepackage{wasysym}
\usepackage{booktabs}
\usepackage{amssymb}
\usepackage{amsbsy}
\usepackage{verbatim}
\usepackage{graphicx}
\usepackage{epstopdf}
\usepackage{color}
\usepackage{sidecap}
\usepackage{bm}% bold math
\usepackage{tikz}

\usepackage[colorlinks,bookmarks=false,citecolor=blue,linkcolor=red,urlcolor=blue]{hyperref}

\newcommand*\circled[1]{\tikz[baseline=(char.base)]{
            \node[shape=circle,draw,inner sep=2pt] (char) {#1};}}

\usepackage{cite}

\catcode`@=11 % If we need private TeX macros
\renewcommand\tableofcontents{%
  \section*{\contentsname}%
  \@starttoc{toc}%
}
\catcode`@=12% '@' is no more a character

\def\be{\begin{equation}}
\def\ee{\end{equation}}

\begin{document}

\setlength{\parindent}{0pt}

\title{The quench action approach in finite integrable spin chains}

\author{Vincenzo Alba and Pasquale Calabrese }
\address{$^1$\,SISSA and INFN, via Bonomea 265, 34136 Trieste, Italy. }
%\address{$^2$\,The Rudolf Peierls Centre for Theoretical Physics,
   % Oxford University, Oxford, OX1 3NP, United Kingdom.   }

\date{\today}

%%%%%%%%%%%%%%%%%%%%%%%%%%%%%%%%%%%%%%%%%%%%%%%%%%%%%%%%%%%%%%%%%%%%%%%%%%
\begin{abstract} 

We consider the problem of constructing the stationary state following a quantum quench, 
using the exact overlaps for finite size integrable models. 
We focus on the isotropic Heisenberg spin chain with initial state N\'eel or Majumdar-Ghosh (dimer),
although the proposed approach is valid for an arbitrary integrable model. 
We consider only eigenstates which do not contain zero-momentum strings because the latter 
are affected by fictitious singularities that are very difficult to take into account. 
We show that the fraction of eigenstates that do not contain zero-momentum strings is vanishing 
in the thermodynamic limit. Consequently, restricting to this part of the Hilbert space leads to 
vanishing expectation values of local observables. However, it is possible to reconstruct the 
asymptotic values by properly reweighting the expectations in the considered subspace, at the 
price of introducing finite-size corrections. We also develop a Monte Carlo sampling of the 
Hilbert space which allows us to study larger systems. We accurately reconstruct the expectation 
values of the conserved charges and the root distributions in the stationary state, which turn 
out to match the exact thermodynamic results. The proposed method can be implemented even in 
cases in which an analytic thermodynamic solution is not obtainable.

\end{abstract}

\maketitle

%%%%%%%%%%%%%%%%% INTRODUCTION %%%%%%%%%%%%%%%%%%%%%
\section{Introduction}
\label{intro}

Understanding the out-of-equilibrium dynamics in {\it isolated} quantum 
many-body systems is one of the most intriguing research topics in 
contemporary physics, both experimentally~\cite{bloch-2008,greiner-2002,
kinoshita-2006,hofferberth-2007,trotzky-2012,gring-2012,cheneau-2012,
schneider-2012,kunhert-2013,langen-2013,meinert-2013,fukuhara-2013,
ronzheimer-2013,braun-2014,langen-2015} and theoretically~\cite{polkovnikov-2011,efg-15}. 
The most investigated protocol is that of the {\it quantum quench}, in 
which a system is initially prepared in an eigenstate $|\Psi_0\rangle$ of 
a many-body hamiltonian ${\mathcal H}_0$. Then a global parameter is 
suddenly changed, and the system is let to evolve unitarily under a new 
hamiltonian ${\mathcal H}$. There is now compelling evidence that at long 
times after the quench an equilibrium steady state arises (see for instance 
Ref.~\cite{kinoshita-2006}), although its nature is not fully understood yet. 
In the thermodynamic limit, due to dephasing, it is natural to expect that 
the equilibrium value of any local observable ${\mathcal O}$ is described 
by the so-called diagonal ensemble as 
\begin{equation}
\label{d-ensemble}
\langle{\mathcal O}\rangle_{DE}=\sum\limits_{\lambda}|\langle\Psi_0|\lambda
\rangle|^2\langle\lambda|{\mathcal O}|\lambda\rangle,
\end{equation}
where the sum is over the eigenstates $|\lambda\rangle$ of the post-quench 
hamiltonian ${\mathcal H}$. Moreover, for generic, i.e., non-integrable, models 
the Eigenstate Thermalization Hypothesis~\cite{deutsch-1991,srednicki-1994} 
(ETH) implies that the diagonal ensemble~\eref{d-ensemble} becomes equivalent 
to the usual Gibbs (thermal) ensemble (a conjecture which has been investigated 
in many numerical studies \cite{tvar,alba-2015a}). 

In integrable models, however, the presence of local or 
quasi-local integrals of motion strongly affects the dynamics, preventing the 
onset of thermal behavior. It has been suggested in Ref.~\cite{rigol-2007,
rigol-2008} that in this situation the post-quench steady-state can be described 
by the so-called Generalized Gibbs Ensemble ($GGE$) as 
\begin{equation}
\label{gge}
\langle{\mathcal O}\rangle_{GGE}=\frac{1}{Z}\Tr({\mathcal O}\rho^{GGE}), 
\quad\textrm{with}\quad\rho^{GGE}\equiv\frac{1}{Z}\exp\Big(-\sum_j\beta_jQ_j\Big). 
\end{equation}
Here $Q_j$ are mutually commuting conserved local (and quasi-local) charges, i.e., $[{\mathcal H},Q_j]=0\,
\forall j$ and $[Q_j,Q_k]=0\,\forall j,k$, whereas $\beta_j$ are Lagrange multipliers 
to be fixed by imposing that  $\langle Q_j\rangle_{GGE}=\langle\Psi_0|Q_j|\Psi_0\rangle$. 
This also provides  useful sum rules that will be of interest in this paper. 
The validity of the GGE has been confirmed in non-interacting theories~\cite{calabrese-2006,
cramer-2008,barthel-2008,rossini-2009,calabrese-2011,cazalilla-2012,se-12,mossel-2012a,
collura-2013,fagotti-2013,kcc14,kcc14a,sotiriadis-2014,bkc-14}, whereas in interacting 
ones the scenario is still not fully settled and some recent works~\cite{ilievski-2015a,
cardy-2015} suggest that the $GGE$ description of the steady state  is complete provided 
that the so-called quasi-local charges~\cite{prosen-2014,pereira-2014,ilievski-2015,essler-2015} 
are included in the GGE~\eref{gge}. 

Valuable insights into this issue have been provided by the so-called quench-action 
approach~\cite{caux-2013}. For Bethe ansatz integrable models this method allows one 
to construct the diagonal ensemble~\eref{d-ensemble} directly in the thermodynamic 
limit, provided that the overlaps $\langle\lambda|\Psi_0\rangle$ are known. The 
physical idea is that for large system sizes it is possible to approximate the sum 
over the eigenstates in~\eref{d-ensemble} using a saddle point argument. Specifically, 
one starts with rewriting~\eref{d-ensemble} as 
\begin{equation}
\label{de1}
\langle{\mathcal O}\rangle_{DE}= \sum\limits_{\lambda}\rho^{DE}\langle\lambda|
{\mathcal O}|\lambda\rangle,\quad\textrm{with}\quad\rho^{DE}=\exp(-2{\rm Re}({\mathcal 
E}(\lambda))), 
\end{equation}
where ${\mathcal E}(\lambda)\equiv-\ln\langle\lambda|\Psi_0\rangle$. 
For typical initial states $|\Psi_0\rangle$,
one has ${\mathcal E}\propto L$, with $L$ the system size. This reflects the 
vanishing of the overlaps as $\langle\Psi_0|\lambda\rangle\propto e^{-c L}$.  
As in the standard Thermodynamic Bethe Ansatz (TBA)~\cite{taka-book}, the 
extensivity of ${\mathcal E}$ suggests that in the thermodynamic limit the 
sum in~\eref{de1} is dominated by a saddle point. Remarkably, for some Bethe 
ansatz solvable models and for simple enough initial states, it is possible to 
determine this saddle point {\it analytically}. This has been done successfully, 
for instance, for the quench from the Bose-Einstein condensate
 in the Lieb-Liniger model ~\cite{de-nardis-2014,ga-15,pce-2015,b-15}, 
 for the quench from some product states in the $XXZ$ spin chain \cite{pozsgay-2014A,wouters-2014A}, 
for transport in spin chains \cite{de_luca}, and for some interacting field theories \cite{bse-14}.  
The quench action approach allows in principle also to reconstruct the full relaxation dynamics 
to the steady state \cite{caux-2013} as numerically done for the 1D Bose 
gas~\cite{de-nardis-2015a,de-nardis-2015}. 

\paragraph*{Outline of the results.} 
In this paper, by combining exact Bethe ansatz and Monte Carlo techniques, we investigate 
the diagonal ensemble and the quench action approach in {\it finite} size integrable models. 
We consider the spin-$1/2$ XXZ spin-chain with $L$ sites, which is defined by the Hamiltonian 
\begin{equation}
\label{xxx-ham}
{\mathcal H}\equiv J\sum\limits_{i=1}^L\left[\frac{1}{2}(S_i^+S^-_{i+1} 
+S_i^{-}S_{i+1}^+)+\Delta S_i^zS_{i+1}^z-\frac{1}{4}\right].
\end{equation}
Here $S^{\pm}_i\equiv (\sigma_i^x\pm i\sigma_i^y)/2$ are spin operators acting on the 
site $i$, $S_i^z\equiv\sigma_i^z/2$, and $\sigma^{x,y,z}_i$ the Pauli matrices. We 
fix $J=1$ in~\eref{xxx-ham} and use periodic boundary conditions, identifying sites 
$L+1$ and $1$. The total magnetization $S_{T}^z\equiv\sum_iS_i^z=L/2-M$, with $M$ number 
of down spins (particles), commutes with~\eref{xxx-ham}, and it is used to label its 
eigenstates. The $XXZ$ chain is Bethe ansatz solvable~\cite{bethe-1931,kor-book} and its eigenstates 
are in one-to-one correspondence with the solutions of the so-called Bethe equations 
(see~\ref{sec:1.2}). 
The non-equilibrium quench dynamics of the XXZ spin chain has been investigated 
numerically intensively \cite{dmcf-06,bar-09,krs-12,cbp-13,cce-15,fagotti-2014}, the GGE with all the 
local charges has been explicitly constructed \cite{fagotti-2014,pozsgay-2013,fagotti-2013b}, 
and the quench action solution has been provided for some initial states~\cite{pozsgay-2014A,wouters-2014A}. 

For concreteness reasons, we focus on the isotropic Heisenberg point ($XXX$), i.e. 
Hamiltonian~\eref{xxx-ham} with $\Delta=1$, but we stress that our approach applies to arbitrary 
$\Delta$, and  also to arbitrary integrable models. We consider the quenches from the zero-momentum 
N\'eel state $|N\rangle$ and the Majumdar-Ghosh state $|MG\rangle$, which are defined as  
\begin{equation}
\label{psi0}
|N\rangle\equiv\Big(\frac{|N_1\rangle
+|N_2\rangle}{\sqrt{2}}\Big), \qquad 
|MG\rangle\equiv \Big(\frac{|\uparrow\downarrow\rangle-|\downarrow\uparrow\rangle}
{\sqrt{2}}\Big)^{\otimes L/2}. 
\end{equation}
Here $|N_1\rangle\equiv |\uparrow\downarrow\rangle^{\otimes L/2}$, and $|N_2\rangle
\equiv |\downarrow\uparrow\rangle^{\otimes L/2}$, with $\otimes$ denoting the tensor 
product. Note that both $|N\rangle$ and $|MG\rangle$ are invariant under one site 
translations. Our study relies on the analytical knowledge of the overlaps between 
the N\'eel and Majumdar-Ghosh state and the eigenstates of~\eref{xxx-ham}~\cite{
brockmann-2014,brockmann-2014b,brockmann-2014c,pozsgay-2014a,piroli-2014,mazza-2015}. 

We first present a detailed overview of the distribution of the overlaps between 
the initial states and the eigenstates of the model (overlap distribution function). 
Precisely, we provide numerical results for all the overlaps for finite chains up 
to $L\le 38$. Our results are obtained exploiting the analytical formulas 
presented in Ref.~\cite{brockmann-2014} and~\cite{pozsgay-2014a}. Crucially, 
we restrict ourselves to a truncated Hilbert space, considering only eigenstates 
of~\eref{xxx-ham} that do not contain zero-momentum strings. Physically, 
zero-momentum strings correspond to eigenstates amplitudes containing multi-particle  
bound states having zero propagation velocity. From the Bethe ansatz perspective, 
in the thermodynamic limit the presence of zero-momentum strings leads to fictitious 
singularities in the overlap formulas. Dealing with these singularities is 
a formidable task that requires detailed knowledge of the solutions of the 
Bethe equations, and it can be done only in very simple cases, e.g., for small 
chains (as explicitly done for the attractive Lieb-Liniger gas for small number 
of particles \cite{calabrese-2014}). On the other hand, it has been argued that 
zero-momentum strings are irrelevant for the reconstruction of the representative 
state in the thermodynamic limit~\cite{wouters-2014A}. 

For a finite chain, we find that for both the N\'eel and the Majumdar-Ghosh initial states, 
the fraction of eigenstates that do not contain zero-momentum strings 
is vanishing in the thermodynamic limit, meaning that for finite large-enough 
systems the vast majority of the eigenstates contain zero momentum strings. Specifically, 
the total number of eigenstates without zero-momentum strings is given in terms of the 
chain size $L$ by simple combinatorial formulas that we provide. 
We investigate the effect of the Hilbert space truncation on the diagonal ensemble focusing on the  
sum rules $\langle Q_j\rangle_{DE}=\langle\Psi_0|Q_j|\Psi_0\rangle$ for the conserved quantities $Q_j$. 
We numerically demonstrate that truncating the diagonal ensemble to 
eigenstates with no zero-momentum strings leads to striking violations of the sum rules,  
as one would naively expect since most of the states with non zero overlaps have not been included. 
Precisely, we numerically observe that the truncated diagonal ensemble 
average $\langle Q_j\rangle$ vanish in the thermodynamic limit reflecting the 
vanishing behavior of the fraction of eigenstates with finite-momentum strings. 
Thus, for finite chains, eigenstates corresponding 
to zero-momentum strings cannot be trivially neglected when considering 
diagonal ensemble averages. 

However, this does not imply that the diagonal ensemble cannot be reconstructed  
from the eigenstates without zero-momentum strings and it is not in contrast 
with the previous quench action results in the thermodynamic limit \cite{wouters-2014A,pozsgay-2014A}.
Indeed, we show  that the correct diagonal ensemble 
sum rules can be recovered in the thermodynamic limit by appropriately 
reweighting the contribution of the finite-momentum strings eigenstates.  

We also develop a Monte Carlo scheme, which is a generalization of 
the approach presented in Ref.~\cite{alba-2015} to simulate the Generalize 
Gibbs Ensemble. The method is based on the knowledge of the N\'eel overlaps 
$\langle\lambda|N\rangle$ and on the knowledge of the Hilbert space structure 
of the $XXX$ chain in the Bethe ansatz formalism. The approach allows us to 
simulate effectively chain with $L\lesssim 60$, although larger systems 
sizes can in principle be reached. 

Strikingly, after the reweighting, although for small chains violations of 
the conserved quantities sum rules are present, these violations vanish in the 
thermodynamic limit and the sum rules are restored. This implies that the only 
effect of the Hilbert space truncation is to introduce finite size scaling corrections. 
In the quench action language, this means that the eigenstates corresponding to non 
zero-momentum strings contain enough physical information about the saddle point. 
This is numerically confirmed by extracting the so-called saddle point 
root distributions, which in the Bethe ansatz language fully characterize 
the diagonal ensemble averages in the thermodynamic limit. In the numerical  
 approach these are obtained from the histograms of the Bethe ansatz 
solutions sampled during the Monte Carlo. Apart from finite size scaling 
corrections, we observe striking agreement with analytical results, at least 
for the first few root distributions.

%%%%%%%%%%%%%%%%%%%%%%% BETHE ANSATZ FOR THE XXX CHAIN %%%%%%%%%%%%%%%%%%%%%%%%%%
\section{Bethe ansatz solution of the Heisenberg ($XXX$) spin chain}
\label{sec:1}

In this section we review some Bethe ansatz results for the spin-$\frac{1}{2}$ Heisenberg 
($XXX$) chain. Specifically, in subsection~\ref{sec:1.2} we discuss the structure of  
its eigenstates (Bethe states) and the associated Bethe equations. 
Subsection~\ref{sec:1.3} focuses on the string hypothesis 
and the so-called Bethe-Gaudin-Takahashi (BGT) equations. 
The form of the BGT equations in the thermodynamic limit is discussed in subsection~\ref{sec:1.4}. 
In subsection~\ref{app-1} we provide some combinatorial formulas for the total 
number of the so-called parity-invariant eigenstates. The latter are the only eigenstates 
having non-zero overlap with the N\'eel and Majumdar-Ghosh states. 
Finally, in subsection~\ref{sec:1.5} we provide the exact formulas for the 
local conserved charges of the model.

%%%%%%%%%%%%%%%%%%%%%%%%%%%%%%%%%%%%%%%%%%%%%%%%%%%%%%%%%%%%%%%%%%%%%%%%%%%
\subsection{Bethe equations and wavefunctions}
\label{sec:1.2}

In the Bethe ansatz framework~\cite{bethe-1931,taka-book} the generic eigenstate 
of~\eref{xxx-ham} (Bethe state) in the sector with $M$ particles can be written as 
\begin{equation}
\label{ba-eig}
|\Psi_M\rangle=\sum\limits_{1\le x_1<x_2<\dots<x_M\le L}A_M(x_1,x_2,
\dots,x_M)|x_1,x_2,\dots,x_M\rangle,
\end{equation}
where the sum is over the positions $\{x_i\}_{i=1}^M$ of the particles, and $A_M(x_1,
x_2,\dots,x_M)$ is the eigenstate amplitude corresponding to the particles 
being at positions $x_1,x_2,\dots, x_M$. The amplitude $A_M(x_1,x_2,\dots, x_M)$ is 
given as 
\begin{equation}
\label{ba_amp}
A_M(x_1,x_2,\dots,x_M)\equiv\sum\limits_{\sigma\in S_M}\exp\Big[i
\sum\limits_{j=1}^Mk_{\sigma_j}x_j+i\sum\limits_{i<j}\theta_{\sigma_i,\sigma_j}
\Big], 
\end{equation}
where the outermost summation is over the permutations $S_M$ of the so-called 
quasi-momenta $\{k_\alpha\}_{\alpha=1}^M$. The two-particle scattering phases 
$\theta_{\alpha,\beta}$ are defined as 
\begin{equation}
\label{s_phases}
\theta_{\alpha,\beta}\equiv \frac{1}{2i}\ln\Big[-\frac{e^{ik_\alpha+ik_\beta}-
2e^{ik_\alpha}+1}{e^{ik_\alpha+ik_\beta}-2e^{ik_\beta}+1}\Big].
\end{equation}
The eigenenergy associated with the eigenstate~\eref{ba-eig} is  
\begin{equation}
\label{ba-ener}
E=\sum\limits_{\alpha=1}^M(\cos(k_\alpha)-1). 
\end{equation}
The quasi-momenta $k_\alpha$ are obtained by solving the so-called Bethe 
equations~\cite{bethe-1931}
\begin{equation}
\label{ba-eq}
e^{ik_\alpha L}=\prod\limits^M_{\beta\ne\alpha}\Big[-\frac{1-2e^{
ik_\alpha}-e^{ik_\alpha+ik_\beta}}{1-2e^{ik_\beta}-e^{ik_\alpha+
ik_\beta}}\Big].
\end{equation}
It is useful to  introduce the rapidities $\{\lambda_\alpha\}_{\alpha=1}^M$ as 
\begin{equation}
\label{rap}
k_\alpha=\pi-2\arctan(\lambda_\alpha)\quad\mbox{mod}\, 2\pi.
\end{equation}
Taking the logarithm on both sides in~\eref{ba-eq} and using~\eref{rap}, 
one obtains the Bethe equations in logarithmic form as 
\begin{equation}
\label{ba-eq-log}
\arctan(\lambda_\alpha)=\frac{\pi}{L}J_\alpha+\frac{1}{L}\sum\limits_{
\beta\ne\alpha}\arctan\Big(\frac{\lambda_\alpha-\lambda_\beta}{2}\Big),
\end{equation}
where $-L/2<J_\alpha\le L/2$ are the so-called Bethe quantum numbers. It can 
be shown that $J_\alpha$ is half-integer(integer) for $L-M$ even(odd)~\cite{taka-book}. 

Importantly, the $M$-particle Bethe states~\eref{ba-eig} corresponding to 
{\it finite} rapidities are eigenstates with maximum allowed magnetization 
(highest-weight eigenstates) $S_T^z=L/2-M=S_T$, with $S_T$ the total spin. 
Due to the $SU(2)$ invariance of~\eref{xxx-ham}, all the states in the same 
$S_T$ multiplet and with different $-S_T\le S_T^z\le S_T$ are eigenstates of the 
$XXX$ chain, with the same energy eigenvalue. These eigenstates 
(descendants) are obtained by multiple applications of the total-spin lowering 
operator $S_T^-\equiv\sum_iS_i^-$ onto the highest-weight states. In the Bethe 
ansatz framework, given a highest-weight eigenstate with $M'$ particles (i.e., 
$M'$ finite rapidities), its descendants are obtained by supplementing the $M'$ 
rapidities with infinite ones. We anticipate that descendant eigenstates 
are important here since they have non-zero overlap with the N\'eel state (cf. 
section~\ref{sec:2}). 
 
%%%%%%%%%%%%%%%%%%%%%%%%%%%%%%%%%%%%%%%%%%%%%%%%%%%%%%%%%%%%%%%%%%%%%%%%%%%
\subsection{String hypothesis \& the Bethe-Gaudin-Takahashi (BGT) equations}
\label{sec:1.3}

In the thermodynamic limit $L\to\infty$ the solutions of the Bethe equations~\eref{ba-eq} 
form particular ``string'' patterns in the complex plane, (string hypothesis)~\cite{
bethe-1931,taka-book}. Specifically, the rapidities forming a ``string'' of length $1
\le n\le M$ (that we defined here as $n$-string) can be parametrized as 
\begin{equation}
\label{str-hyp}
\lambda^{j}_{n;\gamma}=\lambda_{n;\gamma}-i(n-1-2j)+i\delta_{n;\gamma}^j,\qquad 
j=0,1,\dots, n-1, 
\end{equation}
with $\lambda_{n;\gamma}$ being the real part of the string (string center), 
$\gamma$ labelling strings with different centers, and $j$ labelling the different 
components of the string. In~\eref{str-hyp} $\delta_{n;\gamma}^j$ are the string 
deviations, which typically, i.e., for most of the chain eigenstates, vanish 
exponentially with $L$ in the thermodynamic limit. A notable execption are the 
zero-momentum strings for which string devitions exhibit power-law decay. Note that real rapidities 
correspond to strings of unit length ($1$-strings, i.e., $n=1$ in~\eref{str-hyp}). 

The string centers $\lambda_{n;\gamma}$ are obtained by solving the so-called 
Bethe-Gaudin-Takahashi equations~\cite{taka-book}
\begin{equation}
\label{bgt-eq}
2L\theta_n(\lambda_{n;\gamma})=2\pi I_{n;\gamma}+\sum\limits_{(m,
\beta)\ne(n,\gamma)}\Theta_{m,n}(\lambda_{n;\gamma}-\lambda_{m;\beta}).  
\end{equation}
Here the generalized scattering phases $\Theta_{m,n}(x)$ read 
\begin{eqnarray}
\label{Theta}
\nonumber\fl\Theta_{m,n}(x)\equiv\left\{\begin{array}{cc}
\theta_{|n-m|}(x)+\!\!\!\!\!\sum
\limits_{r=1}^{(n+m-|n-m|-1)/2}\!\!\!\!\!2\theta_{|n-m|+2r}(x)
+\theta_{n+m}(x) & \quad\mbox{if}
\quad n\ne m\\\fl\sum\limits_{r=1}^{n-1}2\theta_{2r}(x)+
\theta_{2n}(x) & \quad\mbox{if}\quad n=m
\end{array}\right.
\end{eqnarray}
with $\theta_\alpha(x)\equiv 2\arctan(x/\alpha)$, and $I_{n;\gamma}$  the 
Bethe-Takahashi quantum numbers associated with $\lambda_{n;\gamma}$. 
The solutions of~\eref{bgt-eq}, and the Bethe states~\eref{ba-eig} thereof, 
are naturally classified according to their ``string content'' ${\mathcal S}
\equiv\{s_n\}_{n=1}^M$, with $s_n$ the number of $n$-strings. Clearly, the 
constraint $\sum_{n=1}^{M}n s_n=M$ has to be satisfied. It can be shown that 
the BGT quantum numbers $I_{n;\gamma}$ associated with the $n$-strings are 
integers and half-integers for $L-s_n$ odd and even, respectively. 
Moreover, an upper bound for $I_{n;\gamma}$ can be derived as~\cite{taka-book} 
\begin{equation}
|I_{n;\gamma}|\le I^{(MAX)}_{n}\equiv\frac{1}{2}(L-1-\sum
\limits_{m=1}^Mt_{m,n}s_m),
\label{bt-qn-bound}
\end{equation}
where $t_{m,n}\equiv 2\mbox{min}(n,m)-\delta_{m,n}$. Using the string 
hypothesis~\eref{str-hyp} the Bethe states energy eigenvalue~\eref{ba-ener} 
becomes
\begin{equation}
\label{ener-str}
E=-\sum_{n,\gamma}\frac{2n}{\lambda_{n;\gamma}^2+n^2}. 
\end{equation}
%

%%%%%%%%%%%%%%%%%%%%%%%%%%%%%%%%%%%%%%%%%%%%%%%%%%%%%%%%%%%%%%%%%%%%%%%%%%%
\subsection{The Thermodynamic Bethe Ansatz}
\label{sec:1.4}

In the thermodynamic limit $L\to\infty$ at fixed finite particle density $M/L$ 
the roots of the BGT equations~\eref{bgt-eq} become dense. One then defines 
the BGT root distributions for the $n$-strings as $\pmb{\rho}\equiv\{\rho_n(
\lambda)\}_{n=1}^\infty$, with $\rho_n(\lambda)\equiv\lim_{L\to\infty}[
\lambda_{n;\gamma+1}-\lambda_{n;\gamma}]^{-1}$. Consequently, the  BGT 
equations~\eref{bgt-eq} become an infinite set of coupled non-linear integral 
equations for the $\rho_n(\lambda)$ as 
\begin{equation}
\label{bgt-th}
a_n(\lambda)=\rho_n(\lambda)+\rho^h_n(\lambda)+\sum_m(T_{n,m}*\rho_m)
(\lambda),
\end{equation}
where $\rho_n^{h}(\lambda)$ are the so-called hole-distributions, and the functions 
$a_n(\lambda)$ are defined as 
\begin{equation}
a_n(x)\equiv\frac{1}{\pi}\frac{n}{x^2+n^2}. 
\end{equation}
In~\eref{bgt-th} $T_{n,m}*\rho_m$ denotes the convolution 
\begin{equation}
(T_{n,m}*\rho_m)(\lambda)\equiv\int_{-\infty}^{+\infty}T_{n,m}(\lambda-\lambda')
\rho_{m}(\lambda'),
\end{equation}
with the matrix $T_{n,m}(x)\equiv\Theta'(x)$ being defined as 
\begin{eqnarray}
\nonumber\fl T_{m,n}(x)\equiv\left\{\begin{array}{cc}
a_{|n-m|}(x)+\!\!\!\!\!\sum
\limits_{r=1}^{(n+m-|n-m|-1)/2}\!\!\!\!\!2a_{|n-m|+2r}(x)
+a_{n+m}(x) & \quad\mbox{if}
\quad n\ne m\\\fl\sum\limits_{r=1}^{n-1}2a_{2r}(x)+
a_{2n}(x) & \quad\mbox{if}\quad n=m
\end{array}\right.
\end{eqnarray}
Given a generic, smooth enough, observable ${\mathcal O}$, in thermodynamic 
limit its eigenstate expectation value is replaced by a functional of the root 
densities $\pmb{\rho}$ as $\langle\pmb{\rho}|{\mathcal O}|\pmb{\rho}\rangle$. 
Moreover, for all the local observables ${\mathcal O}$ (the ones considered here) 
the contribution of the 
different type of strings factorize, and $\langle\pmb{\rho}|{\mathcal O}|\pmb{\rho}
\rangle$ becomes 
\begin{equation}
\langle\pmb{\rho}|{\mathcal O}|\pmb{\rho}\rangle=\sum_{n=1}^\infty
\int_{-\infty}^{+\infty}d\lambda \rho_n(\lambda) {\mathcal O}_n(\lambda), 
\end{equation}
with ${\mathcal O}_n(\lambda)$ the $n$-string contribution to the expectation 
value of ${\mathcal O}$.

%%%%%%%%%%%%%%%%%%%%%%%%%%%%%%%%%%%%%%%%%%%%%%%%%%%%%%%%%%%%%%%%%%%%%%%%%%%
\subsection{Parity-invariant eigenstates with non-zero N\'eel and Majumdar-Ghosh overlap: 
counting and string content}
\label{app-1}

Here we provide some exact combinatorial formulas for the total number of 
parity-invariant eigenstates of the $XXX$ chain in the sector with $L/2$ 
particles. These are the only eigenstates having, in principle, non-zero 
overlap with the N\'eel state and the Majumdar-Ghosh state~\cite{brockmann-2014}. 
Parity-invariant eigenstates correspond to solutions of the Bethe equations 
containing only pairs of rapidities with opposite sign. In turn, these  
eigenstates are in one-to-one correspondence with parity-invariant BGT quantum number 
configurations. 

For simplicity we restrict ourselves to the situation with $L$ divisible by four. 
The strategy of the proof is the same as that used to count the number of solutions 
of the Bethe-Gaudin-Takahashi equations (see for instance Ref.~\cite{faddeev-1996}). 
Specifically, the idea is to count all the possible BGT quantum numbers configurations 
corresponding to parity-invariant eigenstates (cf. section~\ref{sec:1.3}). 

We anticipate that the total number of parity-invariant eigenstates $Z_{Neel}$ 
for a chain of length $L$ is given as 
\begin{equation}
\label{N-count}
Z_{Neel}=2^{\frac{L}{2}-1}+\frac{1}{2}B\Big(\frac{L}{2},\frac{L}{4}\Big)+1, 
\end{equation}
with $B(x,y)\equiv x!/(y!(x-y)!)$ the Newton binomial. 
On the other hand, after excluding the zero-momentum strings one obtains 
\begin{equation}
\label{N-count-nz}
\widetilde Z_{Neel}=B\Big(\frac{L}{2},\frac{L}{4}\Big). 
\end{equation}
Note that~\eref{N-count} is only an upper bound for the number of eigenstates 
with non-zero N\'eel overlap, while~\eref{N-count-nz} is exact. Before 
proceeding, we should stress that since the N\'eel state is not $SU(2)$ invariant, 
eigenstates with non-zero N\'eel overlap can contain infinite rapidities. Thus, 
one has to consider all the possible sectors with $L/2=\ell+N_{\infty}$, and $\ell$ 
($N_\infty$) the number of finite (infinite) rapidities. Notice that this 
is different for the Majumdar-Ghosh state, for which only the parity-invariant 
eigenstates in the sector with $\ell=L/2$ have to be considered (see below). 

%%%%%%%%%%%%%%%%%%%%%%%%%%%%%%%%%%%%%%%%%%%%%%%%%%%%%%%%%%%%%%%%%%%%%%%%%%%
\subsubsection{\underline{Parity-invariant states with $\ell$ finite rapidities.}}
\label{app-1.1}

Let us first consider the eigenstate sector with fixed number of finite rapidities 
$\ell$, the remaining $L/2-\ell$ ones being infinite (see section~\ref{sec:1.2}). 
Let us denote the associate string content as ${\mathcal S}=\{s_1,s_2,\dots,
s_{\ell}\}$. Here $s_n$ is the number of $n$-strings, with the constraint 
$\sum_k ks_k=\ell$. It is straightforward to check that the total number of 
parity-invariant quantum number pairs ${\mathcal N}_n(L,{\mathcal S})$ in the 
$n$-string sector is given as 
\begin{equation}
\label{NnLS}
{\mathcal N}_n(L,{\mathcal S})=\Big\lfloor\frac{L}{2}-\frac{1}{2}
\sum_{m=1}^{\ell}t_{nm}s_m\Big\rfloor,
\end{equation}
where $t_{nm}\equiv 2\textrm{Min}(n,m)-\delta_{n,m}$. Thus, the number of parity-invariant 
eigenstates of the $XXX$ chain ${\mathcal N}(L,{\mathcal S})$ compatible with string content 
${\mathcal S}$ is obtained by choosing in all the possible ways the associated parity-invariant 
quantum number pairs as     
\begin{equation}
\label{NLS}
{\mathcal N}(L,{\mathcal S})=\prod_{m=1}^{\ell} B\left({\mathcal N}_m,\left\lfloor
\frac{s_m}{2}\right\rfloor\right).
\end{equation}
Here the product is because each string sector is treated independently, while the 
factor $1/2$ in $s_m/2$ is because since all quantum numbers are organized in pairs, 
only half of them have to be specified. Note that in each $n$-string sector only one 
zero momentum (i.e., zero quantum number) string is allowed, due to the fact that 
repeated solutions of the BGT equation are discarded. Moreover, from~\eref{NnLS} one has 
that $s_m$ is odd (even) only if this zero momentum string is (not) present. 

We now proceed to consider the string configurations with fixed particle number $\ell\le 
L/2$ and fixed number of strings $1\le q\le\ell$. Note that due to parity invariance 
$\ell$ must be even. 
Also, in determining $q$ strings of different length are treated equally, i.e., 
$q=\sum_m s_m$. For a given fixed pair 
$\ell,q$ the total number of allowed quantum number configurations by definition is given 
as 
\begin{equation}
\label{NLlq}
{\mathcal N}'(L,\ell,q)=\sum\limits_{\{\{s_m\}\,:\, \sum m s_m=\ell, \sum s_m=q\}}
{\mathcal N}(L,{\mathcal S}),
\end{equation}
where the sum is over the string content $\{s_m\}_{m=1}^\ell$ compatible with the constraints 
$\sum_m s_m=q$ and $\sum_m m s_m=\ell$. The strategy now is to write a recursive relation 
in both $\ell,q$ for ${\mathcal N}'(L,\ell,q)$. It is useful to consider a shifted string 
content ${\mathcal S}'$ defined as  
\begin{equation}
{\mathcal S}'\equiv \{s_{m+1}\}\quad\textrm{with}\, s_m\in{\mathcal S},\,\forall m.
\end{equation}
Using the definition of $t_{ij}$, it is straightforward to derive that  
\begin{equation}
t_{ij}=t_{i-1,j-1}+2,
\end{equation}
which implies that ${\mathcal N}_n(L,{\mathcal S})$ (see~\eref{NnLS}) satisfies the 
recursive equation 
\begin{equation}
\label{inter}
{\mathcal N}_n(L,{\mathcal S})={\mathcal N}_{n-1}(L-2q,{\mathcal S}'). 
\end{equation}
After substituting~\eref{inter} in~\eref{NLS} one obtains 
\begin{equation}
\label{NLSr}
{\mathcal N}(L,{\mathcal S})=B\Big({\mathcal N}_1(L,{\mathcal S}),\Big\lfloor 
\frac{s_1}{2}\Big\rfloor\Big){\mathcal N}(L-2q,{\mathcal S}'). 
\end{equation}
Finally, using~\eref{NLSr} in~\eref{NLlq}, one obtains a recursive relation for 
${\mathcal N}'(L,\ell,q)$ as 
\begin{equation}
\label{NpLlq}
{\mathcal N}'(L,\ell,q)=\sum_{s=0}^{q-1}B\Big(\frac{L}{2}-q+\Big\lfloor
\frac{s}{2}\Big\rfloor,\left\lfloor\frac{s}{2}\right\rfloor\Big){\mathcal N}'
\left(L-2q,\ell-q,q-s\right), 
\end{equation}
with the condition that for $\ell=q$ one has 
\begin{equation}
{\mathcal N}'(L,q,q)=B\Big(\Big\lfloor\frac{L-q}{2}\Big\rfloor,\Big\lfloor
\frac{q}{2} \Big\rfloor\Big).
\end{equation}
This is obtained by observing that if $\ell=q$ only $1$-strings are allowed 
and~\eref{NnLS} gives ${\mathcal N}_n(L,{\mathcal S})=\lfloor (L-q)/2\rfloor$. 
It is straightforward to check that to satisfy~\eref{NpLlq} for $q$ even one has 
to choose 
\begin{equation}
\label{inter1}
{\mathcal N}'(L,\ell,q)=\frac{q}{\ell}B\Big(\frac{L-\ell}{2},\frac{q}{2}\Big)
B\Big(\frac{\ell}{2},\frac{q}{2}\Big),
\end{equation}
Instead, for $q$ odd one has
\begin{equation}
\label{inter2}
{\mathcal N}'(L,\ell,q)=\frac{\ell-q+1}{\ell}B\Big(\frac{L-\ell}{2},\frac{q-1}{2}
\Big)B\Big(\frac{\ell}{2},\frac{q-1}{2}\Big).
\end{equation}
The number of eigenstates in the sector with $\ell$ particles having nonzero N\'eel 
overlap $Z'_{Neel}(L,\ell)$ is obtained from~\eref{inter1} and~\eref{inter2} by 
summing over all possible values of $q$ as 
\begin{equation}
\label{sum1}
Z'_{Neel}(L,\ell)=\sum\limits_{q=1}^\ell {\mathcal N}'(L,\ell,q).
\end{equation}
It is convenient to split the summation in~\eref{sum1} considering odd and even 
$q$ separately. For $q$ odd one obtains 
\begin{equation}
\sum\limits_{k=0}^{\ell/2-1} {\mathcal N}'(L,\ell,2k+1)=B\Big(\frac{L}{2}-1,
\frac{\ell}{2}-1\Big),
\end{equation}
while for $q$ even one has 
\begin{equation}
\sum\limits_{k=0}^{\ell/2} {\mathcal N}'(L,\ell,2k)=B\Big(\frac{L}{2}-1,
\frac{\ell}{2}\Big). 
\end{equation}
Putting everything together one obtains 
\begin{equation}
\label{N-count-p}
Z'_{Neel}(L,\ell)=B\Big(\frac{L}{2}-1,
\frac{\ell}{2}-1\Big)+B\Big(\frac{L}{2}-1,
\frac{\ell}{2}\Big). 
\end{equation}
The total number of eigenstates with nonzero N\'eel overlap $Z_{Neel}(L)$ 
(cf.~\eref{N-count}) is obtained from~\eref{N-count-p} by summing over the allowed 
values of $\ell=2k$ with $k=0,1,\dots,L/2$. Note that the sum is over $\ell$ even 
due to the parity invariance. 

Finally, it is interesting to observe that the total number $Z_{MG}$ of parity-invariant 
eigenstates having non zero overlap with the Majumdar-Ghosh state is obtained from 
Eq~\eref{N-count-p} by replacing $\ell=L/2$, to obtain 
\begin{equation}
\label{p-inv-mg}
Z_{MG}=B\Big(\frac{L}{2}-1,\frac{L}{4}-1\Big)+B\Big(\frac{L}{2}-1,\frac{L}{4}
\Big). 
\end{equation}
Physically, this is due to the fact that the Majumdar-Ghosh state is invariant under 
$SU(2)$ rotations, implying that only eigenstates with zero total spin $S=0$ can have 
non-zero overlap. 

%%%%%%%%%%%%%%%%%%%%%%%%%%%%%%%%%%%%%%%%%%%%%%%%%%%%%%%%%%%%%%%%%%%%%%%%%%%
\subsubsection{\underline{Excluding the zero-momentum strings.}}
\label{app-1.2}

One should first observe that for the generic eigenstate of the $XXX$ chain with 
$\ell$ finite rapidities, due to parity invariance and the exclusion of zero-momentum 
strings, only $n$-strings with 
length $n\le \ell/2$ are allowed. Also, the string content can be written as 
${\mathcal S}\equiv\{s_1,\dots,s_{\ell/2}\}$, i.e., $s_m=0$ $\forall m>
\ell/2$. Due to the parity invariance one has that $s_m$ is always even. Clearly 
one has $\sum_{m=1}^{\ell/2}m s_m=\ell$. Finally, the total number of 
parity-invariant quantum numbers $\widetilde{\mathcal N}_n$ in the $n$-string 
sector is given as  
\begin{equation}
\widetilde{\mathcal N}_n(L,{\mathcal S})=\frac{L}{2}-\frac{1}{2}
\sum_{m=1}^{\ell/2}t_{nm} s_m.
\end{equation}
The proof now proceeds as in the previous section. One defines the total number of eigenstates 
with nonzero N\'eel overlap in the sector with fixed $\ell$ finite rapidities and $q$ different 
string types as $\widetilde{\mathcal N}'(L,\ell,q)$. It is straightforward to show that 
$\widetilde{\mathcal N}'(L,\ell,q)$ obeys the recursive relation
\begin{equation}
\label{NpLlq-1}
\widetilde{\mathcal N}'(L,\ell,q)=\sum_{s=0}^{q/2-1}B\Big(\frac{L}{2}-q+s,s\Big)\widetilde
{\mathcal N}'\Big(L-2q,\frac{\ell-q}{2},\frac{q}{2}-s\Big),
\end{equation}
with the constraint
\begin{equation}
\widetilde{\mathcal N}'(L,1,1)=\frac{L}{2}-1. 
\end{equation}
The solution of~\eref{NpLlq-1} is given as 
\begin{equation}
\widetilde{\mathcal N}'(L,\ell,q)=\frac{L-2\ell+2}{L-\ell+2}B\Big(\frac{L-\ell}{2}+1,q\Big)
B\Big(\frac{\ell}{2}-1,\frac{q}{2}-1\Big).
\end{equation}
After summing over the allowed values of $q=2k$ with $k=1,2,\dots,\ell/2$ one obtains 
the total number of eigenstates with nonzero N\'eel overlap at fixed number of 
particles $\ell$ $\widetilde Z_{Neel}'(L,\ell)$ as 
\begin{equation}
\label{Neel-fi}
\widetilde Z_{Neel}'(L,\ell)=B\Big(\frac{L}{2},\frac{\ell}{2}\Big)-
B\Big(\frac{L}{2},\frac{\ell}{2}-1\Big).
\end{equation}
by summing over $\ell$ one obtains~\eref{N-count-nz}. Similar to~\eref{p-inv-mg} the 
total number of eigenstates $\widetilde Z_{MG}$ having non-zero overlap with the 
Majumdar-Ghosh state is obtained from~\eref{Neel-fi} by replacing $\ell\to L/2$, to obtain 
\begin{equation}
\label{mg-fi}
\widetilde Z_{MG}=B\Big(\frac{L}{2},\frac{L}{4}\Big)-B\Big(\frac{L}{2},\frac{L}{4}-1
\Big). 
\end{equation}
Interestingly, using~\eref{p-inv-mg} and~\eref{mg-fi}, one obtains that the ratio 
$\widetilde Z_{MG}/Z_{MG}$ is given as 
\begin{equation}
\frac{\widetilde Z_{MG}}{Z_{MG}}=\frac{4}{4+L}. 
\end{equation}
%

%%%%%%%%%%%%%%%%%%%%%%%%%%%%%%%%%%%%%%%%%%%%%%%%%%%%%%%%%%%%%%%%%%%%%%%%%%%
\subsection{The conserved charges}
\label{sec:1.5}

The $XXX$ chain exhibits an extensive number of mutually commuting local conserved 
charges~\cite{grabowski-1995} $Q_n$ ($n\in\mathbb{N}$), i.e., 
\begin{equation}
[Q_n,{\mathcal H}]=0\quad\forall n\quad\textrm{and}\quad [Q_n,Q_m]=0\quad\forall n,m.
\end{equation}
The corresponding charges eigenvalues are given as 
\begin{equation}
\label{Q-def}
\left.Q_{n+1}\equiv\frac{i}{(n-1)!}\frac{d^n}{dy^n}\ln\tau
(y)\right|_{y=i},
\end{equation}
where $y$ is a spectral parameter and $\tau(y)$ is the eigenvalue of the so-called 
transfer matrix in the Algebraic Bethe Ansatz framework~\cite{kor-book}. The analytic 
expression for $\tau(y)$ in terms of the solutions $\{\lambda_\alpha\}$ of the Bethe 
equations~\eref{ba-eq} is given as 
\begin{equation}
\label{tau}
\tau(y)\equiv\Big(\frac{y+i}{2}\Big)^L\prod\limits_\alpha\frac{y-\lambda_\alpha-2i}
{y-\lambda_\alpha}+\Big(\frac{y-i}{2}\Big)^L\prod\limits_\alpha\frac{y-\lambda_\alpha
+2i}{y-\lambda_\alpha}.
\end{equation}
Interestingly, one can check that the second term in~\eref{tau} does not contribute to 
$Q_n$, at least for small enough $n\ll L$. For a generic Bethe state, using the string 
hypothesis~\eref{str-hyp} the eigenvalue of $Q_n$ is obtained by summing 
independently the contributions of the BGT roots (see~\eref{bgt-eq}) as 
\begin{equation}
\label{qngnk}
Q_n=\sum_{k,\gamma}g_{n,k}(\lambda_{k;\gamma}).
\end{equation}
Using the string hypothesis (cf.~\eref{str-hyp}) and~\eref{Q-def}~\eref{tau}, one obtains  
the first few functions $g_{n,k}$ in terms of the solutions of the BGT equations~\eref{bgt-eq} 
as 
\begin{eqnarray}
\label{gnk}
g_{2,k}=-\frac{2k}{\lambda^2_{k;\gamma}+k^2}, &\quad g_{3,k}=-\frac{4k\lambda_{k;\gamma}}
{(\lambda_{k;\gamma}^2+k^2)^2}\\\nonumber  
g_{4,k}=\frac{2k(k^2-3\lambda_{k;\gamma}^2)}{(k^2+\lambda_{k;\gamma}^2)^3}, &\quad 
g_{5,k}=\frac{8k\lambda_{k;\gamma}(k^2-\lambda_{k;\gamma}^2)}{(k^2+
\lambda_{k;\gamma}^2)^4}\\\nonumber
g_{6,k}=-\frac{2k(5\lambda_{k;\gamma}^4-10k^2\lambda_{k;\gamma}^2+k^4)}{(k^2+
\lambda_{k;\gamma}^2)^5}. 
\end{eqnarray}
It is interesting to observe that $g_{n,k}$ is vanishing in the limit $\lambda_{k;\gamma}\to\infty$. 
This is expected to hold for the generic $n,k$, and it is a consequence of the $SU(2)$ invariance 
of the conserved charges. Finally, in the thermodynamic limit $L\to\infty$ one can replace the 
sum over $\gamma$ in~\eref{qngnk} with an integral to obtain  
\begin{equation}
\label{q0-th}
q_n\to\sum_{k=1}^\infty\int_{-\infty}^{+\infty}d\lambda\rho_k(\lambda)g_{n,k}(\lambda), 
\end{equation}
where the BGT root distributions $\rho_k(\lambda)$ are solutions of the system of integral 
equations~\eref{bgt-th}.

%%%%%%%%%%%%%%%%%%%%%%%%%%%%%%%%%%%%%%%%%%%%%%%%%%%%%%%%%%%%%%%%%%%%%%%%%%%
\section{Overlap between the Bethe states and some simple product states} 
\label{sec:2}

Here we detail the Bethe ansatz results for the overlap of the Bethe states (cf.~\eref{ba-eig}) 
with the zero-momentum (one-site shift invariant) N\'eel state $|N\rangle$ 
and the Majumdar-Ghosh (MG) $|MG\rangle$ state (cf.~\eref{psi0}). In particular, we specialize 
the Bethe ansatz results to the case of eigenstates described by perfect strings. 

%%%%%%%%%%%%%%%%%%%%%%%%%%%%%%%%%%%%%%%%%%%%%%%%%%%%%%%%%%%%%%%%%%%%%%%%%%%
\subsection{N\'eel state overlaps}
\label{sec:2.1}

We start discussing the overlaps with the N\'eel state. 
As shown in Ref. \cite{brockmann-2014,wouters-2014A}, only parity-invariant Bethe states have non-zero overlap with the N\'eel state. 
The corresponding solutions of the Bethe equations~\eref{ba-eq} contain only pairs of rapidities with opposite sign.  
Here we denote the generic parity-invariant rapidity configuration as $|\{\pm\tilde\lambda_j\}_{j=1}^m,n_\infty\rangle$, 
i.e., considering only positive rapidities (as stressed by the tilde in $\tilde\lambda_j$). 
Here $m$ is the number of rapidity pairs. Since the N\'eel state is not invariant 
under $SU(2)$ rotations, eigenstates with infinite rapidities can have non-zero N\'eel overlaps.
We denote the number of infinite rapidities as $N_{\infty}$. Note that one has $M=L/2=
N_\infty+2m$. The density of infinite rapidities is denoted as $n_\infty\equiv N_\infty/L$. 
The overlap between the Bethe states and the N\'eel state $|N\rangle$ reads~\cite{wouters-2014A,
pozsgay-2014a} 
\begin{equation}
\label{Neel-ov}
\frac{\langle N|\{\pm\tilde\lambda_j\}_{j=1}^m,n_\infty\rangle}{|||\{\tilde\lambda_j\}_{j=1}^m,
n_\infty\rangle||}=\frac{\sqrt{2}N_{\infty}!}{\sqrt{(2N_\infty)!}}\left[\prod_{j=1}^m
\frac{\sqrt{\tilde\lambda_j^2+1}}{4\tilde\lambda_j}\right]\sqrt{\frac{\textrm{det}_m(G^+)}{
\textrm{det}_m(G^-)}}.
\end{equation}
The matrix $G^\pm$ is  defined as  
\begin{equation}
\label{G-pm}
\fl\quad G^{\pm}_{jk}=\delta_{jk}\Big(LK_{1/2}(\tilde\lambda_j)-\sum\limits_{l=1}^mK_1^+(
\tilde\lambda_j,\tilde\lambda_l)\Big)+K_{1}^{\pm}(\tilde\lambda_j,\tilde\lambda_k),
\quad\,j,k=1,\dots,m, 
\end{equation}
where 
\begin{equation}
\label{K}
K_1^\pm(\lambda,\mu)=K_1(\lambda-\mu)\pm K_1(\lambda+\mu) \quad\textrm{with}\quad 
K_\alpha(\lambda)\equiv\frac{8\alpha}{\lambda^2+4\alpha^2}. 
\end{equation}
Note that our definitions of $K_{\alpha}(\lambda)$ differs from the one in 
Ref.~\cite{brockmann-2014}, due to a factor $2$ in the definition of the 
rapidities (cf.~\eref{str-hyp}). 

%%%%%%%%%%%%%%%%%%%%%%%%%%%%%%%%%%%%%%%%%%%%%%%%%%%%%%%%%%%%%%%%%%%%%%%%%%%
\subsection{The string hypothesis: Reduced formulas for the N\'eel overlaps}
\label{sec:2.2}

Here we consider the overlap formula for the N\'eel state~\eref{Neel-ov} in the 
limit $L\to\infty$, assuming that the rapidities form perfect strings, i.e., 
$\delta_{n;\gamma}^j=0$ in~\eref{str-hyp}. Then it is possible to 
rewrite~\eref{Neel-ov} in terms of the string centers $\tilde\lambda_{n;\alpha}$ 
only. We restrict ourselves to parity-invariant rapidity configurations with no 
zero-momentum strings, i.e., with finite string centers (cf.~\eref{str-hyp}). 
We denote the generic parity-invariant string configuration as $\{\tilde
\lambda_{n;\gamma}\}$, where $\gamma$ labels the different non-zero string 
centers, and $n$ is the string length. Note that due to parity invariance and 
the exclusion of zero-momentum strings, only strings of length up to $m$ are 
allowed. The string 
content (cf.~\ref{sec:1.3}) of parity-invariant Bethe states is denoted as 
$\widetilde{\mathcal S}=\{\tilde s_1,\dots,\tilde s_{m}\}$, with $\tilde s_n$ 
the number of pairs of $n$-strings. 

It is convenient to split the indices $i,j$ in $G^\pm_{ij}$ (cf.~\eref{G-pm}) as 
$i=(n,\gamma,i)$ and $j=(m,\gamma',j)$, with $n,m$ being the length of the strings, 
$\gamma,\gamma'$ labelling the corresponding string centers, and $i,j$ the components 
of the two strings. Using~\eref{G-pm} and~\eref{K}, one has that for two consecutive 
rapidities in the same string, i.e., for $m=n,\gamma=\gamma',|i-j|=2$, the matrices 
$G^{\pm}_{jk}$ become ill-defined in the thermodynamic limit. Precisely, $K_{1}(\tilde
\lambda_{n;\gamma}^i-\tilde\lambda_{n;\gamma}^{i+1})\sim 1/(\delta_{n;\gamma}^i-
\delta_{n;\gamma}^{i+1})$, implying that $G_{ij}^\pm$ diverges in the thermodynamic 
limit. However, as the same type of divergence occurs in both $G^+$ and $G^-$, their 
ratio (cf.~\eref{Neel-ov}) is finite. 

The finite part of the ratio $\det G^+/\det G^-$ can be extracted using the same 
strategy as in Ref.~\cite{calabrese-2007,calabrese-2007-a} (see also Ref.~\cite{brockmann-2014,calabrese-2014}). 
One obtains that $\det G^+/\det G^-\to \det\widetilde G^+/\det\widetilde G^-$. The reduced 
matrix $\widetilde G^+$ depends only on the indices $(n,\gamma)$ 
and $(m,\gamma')$ of the ``string center''  and it is given as  
\begin{equation}
\label{red-G+}
\fl \frac{1}{2}\widetilde G^+_{(n,\gamma)(m,\gamma')}=\left\{\begin{array}{cc}
L\theta_n'(\tilde\lambda_{n;\gamma}) -\sum\limits_{(\ell,\alpha)\ne(n,\gamma)}\Big[\Theta'_{n,\ell}
(\tilde\lambda_{n;\gamma}-\tilde\lambda_{\ell;\alpha}) & \quad\textrm{if}\,(n,\gamma)=(m,\gamma') \\
+\Theta'_{n,\ell}(\tilde\lambda_{n;\gamma}+\tilde\lambda_{\ell;\alpha})\Big] & \\ \\
\Theta'_{n,m}
(\tilde\lambda_{n;\gamma}-\tilde\lambda_{m;\gamma'})+\Theta'_{n,m}
(\tilde\lambda_{n;\gamma}+\tilde\lambda_{m;\gamma'}) & \quad\textrm{if}\,(n,\gamma)\ne(m,\gamma')
\end{array}\right.
\end{equation}
Here $\theta_n'(x)\equiv d\theta_n(x)/dx=2n/(n^2+x^2)$ and $\Theta'(x)\equiv d\Theta(x)/dx$, 
with $\Theta(x)$ as defined in~\eref{Theta}. 
Similarly, for $\widetilde G^-$ one obtains 
\begin{equation}
\label{red-G-}
\fl\frac{1}{2}\widetilde G^-_{(n,\gamma)(m,\gamma')}=\left\{\begin{array}{cc}
\fl(L-1)\theta'_n(\tilde\lambda_{n;\gamma})-2\sum\limits_{k=1}^{n-1}\theta'_k(
\tilde\lambda_{n;\gamma})
& \textrm{if}\,(n,\gamma)= (m,\gamma')\\
-\hspace{-.5cm}\sum\limits_{(\ell,\alpha)\ne(n,\gamma)}\Big[\Theta'_{n,\ell}
(\tilde\lambda_{n;\gamma}-\tilde\lambda_{\ell;\alpha})+\Theta'_{n,\ell}
(\tilde\lambda_{n;\gamma}+\tilde\lambda_{\ell;\alpha})\Big] \\\\
\Theta'_{n,m}
(\tilde\lambda_{n;\gamma}-\tilde\lambda_{m;\gamma'})-\Theta'_{n,m}
(\tilde\lambda_{n;\gamma'}+\tilde\lambda_{m;\gamma'})) & \textrm{if}\,(n,\gamma)\ne(m,\gamma')
\end{array}\right.
\end{equation}
We should stress that in presence of zero-momentum strings, additional divergences as 
$1/(\delta_{n;\gamma}^{i}+\delta_{n;\gamma}^{i+1})$ appear in $\widetilde G^\pm$, due to the term $K_1(\lambda+\mu)$ 
in~\eref{G-pm}. The treatment of these divergences is a challenging task because it requires, for each 
different type of string, the precise knowledge of the string deviations, meaning their dependence 
on $L$. Some results have been provided for small strings in Ref.~\cite{calabrese-2014}. 

Finally, using the string hypothesis and the parity-invariance condition, the prefactor of the 
determinant ratio in~\eref{Neel-ov} becomes  
\begin{equation}
\label{Neel-k}
\fl\quad\prod\limits_{j=1}^m\frac{\sqrt{\tilde\lambda_j^2+1}}{4\tilde\lambda_j}=
\frac{1}{4^m}\prod\limits_{n=1}^m\prod\limits_{\ell=1}^{\tilde s_n}\left[\frac{\sqrt{n^2+
\tilde\lambda^2_{n;\ell}}}{\tilde\lambda_{n;\ell}}
\prod\limits_{k=0}^{\lceil n/2\rceil-1}\frac{(2k)^2+\tilde\lambda^2_{n;\ell}}{(2k+1)^2+
\tilde\lambda^2_{n;\ell}}\right]^{(-1)^n}, 
\end{equation}
where $\tilde s_n$ is the number of $n$-string pairs in the Bethe state.

%%%%%%%%%%%%%%%%%%%%%%%%%%%%%%%%%%%%%%%%%%%%%%%%%%%%%%%%%%%%%%%%%%%%%%%%%%%
\subsection{Overlap with the Majumdar-Ghosh state}
\label{sec:2.3}

The overlap between a generic eigenstate of the $XXX$ chain $|\{\pm\tilde\lambda_j\}
\rangle$ and the Majumdar-Ghosh state can be obtained from the N\'eel state 
overlap~\eref{Neel-ov} as~\cite{pozsgay-2014a} 
\begin{equation}
\label{mg-ov}
\langle MG|\{\pm\tilde\lambda_j\}_{j=1}^m\rangle=\prod\limits_{j=1}^m\frac{1}{2}
\Big(1-\frac{\tilde\lambda_j-i}{\tilde\lambda_j+i}\Big)
\Big(1+\frac{\tilde\lambda_j+i}{\tilde\lambda_j-i}\Big)
\langle N|\{\pm\tilde\lambda_j\}_{j=1}^m\rangle.
\end{equation}
Notice that the Bethe states having non-zero overlap with the Majumdar-Ghosh state 
do not contain infinite rapidities ($N_\infty=0$), in contrast with the N\'eel case 
(cf.~\eref{Neel-ov}). Using the string hypothesis, the multiplicative factor 
in~\eref{mg-ov} is rewritten as 
\begin{eqnarray}
\label{mg-k}
\fl\prod\limits_{j=1}^m\frac{1}{2}
\Big(1-\frac{\tilde\lambda_j-i}{\tilde\lambda_j+i}\Big)
\Big(1+\frac{\tilde\lambda_j+i}{\tilde\lambda_j-i}\Big)=\\\nonumber\qquad
2^m\prod\limits_{n=1}^m\prod\limits_{\ell=1}^{\tilde s_n}
\tilde\lambda_{n;\ell}^{1+(-1)^n}(\tilde\lambda^2_{n;\ell}+n^2)\prod\limits_{k=0}^{
\lfloor n/2\rfloor}\Big[\tilde\lambda_{n;\ell}^2+\Big(2k+\frac{1-(-1)^n}{2}\Big)^2
\Big]^{-2}.
\end{eqnarray}
%

%%%%%%%%%%%%%%%%%%%%%%%%%%%%%%%%%%%%%%%%%%%%%%%%%%%%%%%%%%%%%%%%%%%%%%%%%%%
\subsection{The N\'eel overlap in the thermodynamic limit}
\label{sec:2.4}

In the thermodynamic limit $L\to\infty$ the extensive part of the N\'eel 
overlap~\eref{Neel-ov} can be written as~\cite{brockmann-2014} 
\begin{equation}
\label{Neel-ov-th}
\fl-\lim_{L\to\infty}\ln\left[\frac{\langle N|\{\pm\tilde\lambda_j\}_{j=1}^m,n_\infty
\rangle}{|||\{\tilde\lambda_j\}_{j=1}^m,n_\infty\rangle||}\right]=\frac{L}{2}\Big(
n_\infty\ln2+\sum_{n=1}^\infty\int_0^\infty d\lambda\rho_n(\lambda)[g_n(\lambda)
+2n\ln(4)]\Big), 
\end{equation}
where  
\begin{equation}
\fl\quad g_n(\lambda)=\sum_{l=1}^{n-1}\Big[f_{n-1-2l}(\lambda)-f_{n-2l}(\lambda)
\Big],\quad\textrm{with}\quad f_n(\lambda)=\ln\Big(\lambda^2+\frac{n^2}{4}\Big),
\end{equation}
and
\begin{equation}
n_\infty=1-2\sum_{m=1}^\infty m\int_{-\infty}^\infty d\lambda\rho_m(\lambda). 
\end{equation}
Note that~\eref{Neel-ov-th} is extensive, due to the prefactor $L/2$. Also, 
\eref{Neel-ov-th} is obtained only from~\eref{Neel-k}, while the subextensive 
contributions originating from the determinant ratio $\det_m(G^+)/\det_m(G^-)$ 
in~\eref{Neel-ov} are neglected. We should mention that~\eref{Neel-ov-th} 
acts as a driving term in the quench action formalism (cf. section~\ref{sec:4}).

%%%%%%%%%%%%%%%%%%%%%%%%%%%%%%%%%%%%%%%%%%%%%%%%%%%%%%%%%%%%%%%%%%%%%%%%%%%
\section{Quench action treatment of the steady state}
\label{sec:4}

The quench action formalism~\cite{caux-2013} allows one to construct a saddle 
point approximation for the diagonal ensemble. First, in the 
thermodynamic limit the sum over the chain eigenstates in~\eref{d-ensemble} 
can be recast into a functional integral over the BGT root distributions 
$\pmb{\rho}\equiv\{\rho_n(\lambda)\}_{n=1}^\infty$ (cf. section~\ref{sec:1.4}) 
as
\begin{equation}
\label{eig-sum}
\sum\limits_{\alpha}\rightarrow\int{\mathcal D}\pmb{\rho} e^{S_{YY}(\pmb{\rho})}. 
\end{equation}
Here ${\mathcal D}\pmb{\rho}\equiv\prod_{n=1}^\infty{\mathcal D}\rho_n(\lambda)$
 and $S_{YY}(\pmb{\rho})$ is the Yang-Yang entropy
\begin{equation}
\fl\quad S_{YY}(\pmb{\rho})\equiv L\sum_{n=1}^\infty\int_{-\infty}^{+\infty}d
\lambda\Big[\rho_n(\lambda)\ln\Big(1+\frac{\rho_n^h(\lambda)}{\rho_n(\lambda)}
\Big)+\rho_n^h(\lambda)\ln\Big(1+\frac{\rho_n(\lambda)}{\rho^h_n(\lambda)}\Big)
\Big],
\end{equation}
which counts the number of microscopic Bethe states~\eref{ba-eig} leading to the same 
$\pmb{\rho}$ in the thermodynamic limit. Using~\eref{eig-sum}, the diagonal ensemble 
expectation value~\eref{d-ensemble} of a generic observable ${\mathcal O}$ becomes 
\begin{equation}
\label{qa-d-ensemble}
\quad\langle{\mathcal O}\rangle=\int{\mathcal D}\pmb{\rho}\exp\Big[2\Re\ln\langle
\Psi_0|\pmb{\rho}\rangle +S_{YY}(\pmb{\rho})\Big]\langle\pmb{\rho}|{\mathcal O}|
\pmb{\rho}\rangle.
\end{equation}
Here it is assumed that in the thermodynamic limit the eigenstate expectation 
values $\langle\alpha|{\mathcal O}|\alpha\rangle$ (cf.~\eref{d-ensemble}) 
become smooth functionals of the root distributions $\pmb{\rho}$, whereas 
$\langle\pmb{\rho}|\Psi_0\rangle$ for the N\'eel state is readily obtained 
from~\eref{Neel-ov-th}. 

The functional integral in~\eref{qa-d-ensemble} can be evaluated in the limit 
$L\to\infty$ using the saddle point approximation. One has to minimize the 
functional ${\mathcal F}(\pmb{\rho})$ defined as 
\begin{equation}
\label{qa-term}
L{\mathcal F}(\pmb{\rho})\equiv 2\Re\ln\langle\pmb{\rho}|\Psi_0
\rangle+S_{YY}(\pmb{\rho}(\lambda))  ,
\end{equation}
with respect to $\pmb{\rho}$, i.e., solving $\delta{\mathcal F}(\pmb{\rho})/\delta
\pmb{\rho}|_{\pmb{\rho}=\pmb{\rho^*}}=0$, under the constraint that the thermodynamic BGT 
equations~\eref{bgt-th} hold. Finally, one obtains from~\eref{qa-d-ensemble} that in the 
thermodynamic limit 
\begin{equation}
\label{obs-th}
\langle{\mathcal O}\rangle=\langle\pmb{\rho^*}|{\mathcal O}|\pmb{\rho^*}\rangle. 
\end{equation}
Remarkably, for the quench with initial state the N\'eel state $|\Psi_0\rangle=|N\rangle$ the 
saddle point root distributions ${\rho_n^*(\lambda)}_{n=1}^\infty$ can be obtained 
analytically~\cite{brockmann-2014}. The first few are given as 
\begin{eqnarray}
\label{rho1-sp}
\fl\quad\rho^*_1(\lambda)=\frac{8(4+\lambda^2)}{\pi(19+3\lambda^2)(1+6\lambda^2+
\lambda^4)}, \\\label{rho2-sp}
\fl\quad\rho_2^*(\lambda)=\frac{8\lambda^2(9+\lambda^2)(4+3\lambda^2)}{\pi(2+\lambda^2)
(16+14\lambda^2+\lambda^4)(256+132\lambda^2+9\lambda^4)}, \\\label{rho3-sp}
\fl\quad\rho_3^*(\lambda)=\frac{8(1+\lambda^2)^2(5+\lambda^2)(16+\lambda^2)(21+\lambda^2)}
{\pi(19+3\lambda^2)(9+624\lambda^2+262\lambda^4+32\lambda^6+\lambda^8)
(509+5\lambda^2(26+\lambda^2))}.
\end{eqnarray}
% 

%%%%%%%%%%%%%%%%%%%%%%%%%%%%%%%%%%%%%%%%%%%%%%%%%%%%%%%%%%%%%%%%%%%%%%%%%%%
\section{The role of the zero-momentum strings Bethe states}
\label{sec:5}

In this section we discuss generic features of the overlaps between the 
eigenstates (Bethe states) of the Heisenberg spin chain and the N\'eel state. 
We exploit the Bethe ansatz solution of the chain (see 
section~\ref{sec:1}) as well as exact results for the N\'eel overlaps (see 
section~\ref{sec:2}). We focus on finite chains with $L\lesssim 40$ sites. 
The only Bethe states having non zero N\'eel overlap are the parity-invariant Bethe states (see 
section~\ref{sec:2}). We denote their total number as $Z_{Neel}$. Crucially, 
here we restrict ourselves to the parity-invariant Bethe states that do not 
contain zero-momentum strings. We denote the total number of these eigenstates 
as $\widetilde Z_{Neel}$. 
%Both $Z_{Neel}$ and $\widetilde Z_{Neel}$ are given 
%in terms of the chain length $L$ by simple combinatorial formulas. % that we provide. 

Interestingly, the fraction of eigenstates with no zero-momentum strings, i.e.,  
$\widetilde Z_{Neel}/Z_{Neel}$, is vanishing as $L^{-1/2}$ in the thermodynamic 
limit, meaning that zero-momentum strings eigenstates are dominant in number 
for large chains. This has consequences at the level of the overlap sum rules for the local conservation laws of 
the model that cannot be saturated by a vanishing fraction of states. 
Indeed, limiting ourselves to states that do not contain zero momentum strings, we find  that  all sum rules exhibit vanishing behavior as 
$L^{-1/2}$ upon increasing the chain size, reflecting the same vanishing behavior 
as $\widetilde Z_{Neel}/Z_{Neel}$. 
A similar scenario holds for the overlaps with 
the Majumdar-Ghosh state, where excluding the zero-momentum strings leads to a $1/L$ behavior. 
This shows that zero-momentum strings eigenstates are not naively negligible.

%##################################################################
\begin{figure}[t]
\begin{center}
\includegraphics[width=.9\textwidth]{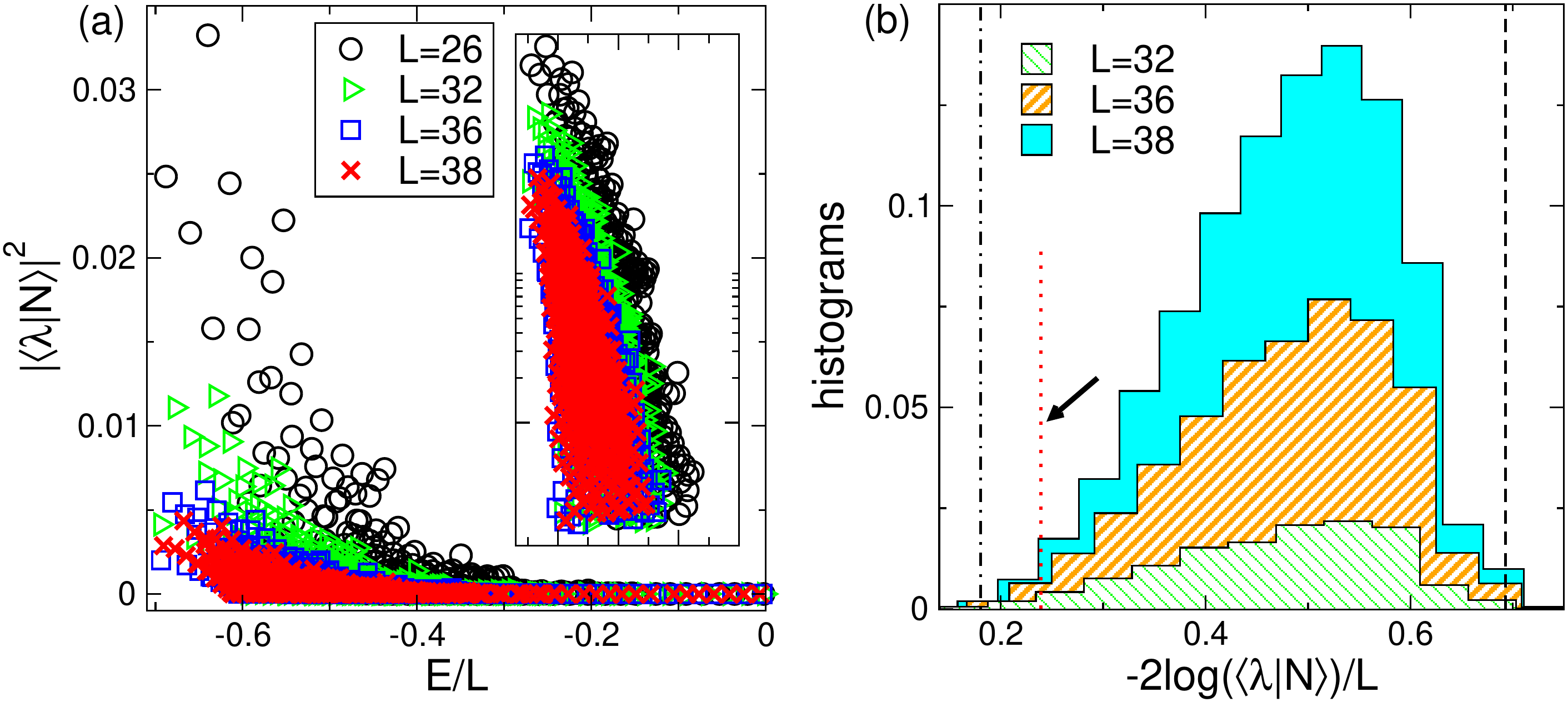}
\end{center}
\caption{ N\'eel overlaps with the eigenstates of the Heisenberg spin 
 chain: Numerical results obtained from the full scanning of the chain 
 Hilbert space. Eigenstates corresponding to zero-momentum strings are 
 excluded. (a) Squared overlaps $|\langle\lambda|N\rangle|^2$ plotted as 
 function of the eigenstates energy density $E/L$. Here $|\lambda\rangle$ 
 denotes the generic eigenstate. The data are for chains with length $26
 \le L\le 38$. The inset highlights the exponential decay as a 
 function of $E/L$ (note the logarithmic scale on the $y$ axis). (b) 
 Overlap distribution function: Histograms of $-2\ln\langle\lambda|N\rangle/L$. 
 The $y$-axis is rescaled by a factor $10^5$ for convenience. The dash-dotted 
 and dashed vertical lines are the N\'eel overlaps with the $XXX$ chain ground 
 state and the ferromagnetic state, respectively. The dotted line (see 
 the arrow) is the result obtained using the quench action approach. 
}
\label{fig0:Neel-ov}
\end{figure}
%##################################################################

%%%%%%%%%%%%%%%%%%%%%%%%%%%%%%%%%%%%%%%%%%%%%%%%%%%%%%%%%%%%%%%%%%%%%%%%%%%
\subsection{N\'eel overlap distribution function: Overview}
\label{sec:5.1}

Here we overview the Bethe ansatz results for the N\'eel overlaps with the 
eigenstates of the $XXX$ chain. The total number of parity-invariant 
eigenstates $Z_{Neel}$ having, in principle, non-zero N\'eel overlap is given 
as 
\begin{equation}
\label{zNeel1}
Z_{Neel}=2^{\frac{L}{2}-1}+\frac{1}{2}B\Big(\frac{L}{2},\frac{L}{4}\Big)+1, 
\end{equation}
with $B(n,m)\equiv n!/(m!(n-m)!)$ the Newton binomial. The proof of~\eref{zNeel1} 
is obtained by counting all the parity-invariant BGT quantum number configurations, 
and it is reported in~\ref{app-1}. Note that $Z_{Neel}$ provides  an upper 
bound for the number of eigenstates with non-zero N\'eel overlap, as it is clear 
from the exact diagonalization results shown in Table~\ref{table:Neel}. This is 
because parity-invariant eigenstates with a single zero-momentum even-length 
string, which are included in~\eref{zNeel1}, have identically zero N\'eel 
overlap~\cite{brockmann-2014}. This is not related to the symmetries of the N\'eel 
state, but to an ``accidental'' vanishing of the prefactor in the overlap 
formula~\eref{Neel-ov}. In particular, this is related to the conservation of quasi-local 
charges~\cite{ilievski-2015a}. Finally, after excluding the zero-momentum strings 
eigenstates, the total number of remaining eigenstates $\widetilde Z_{Neel}$, 
which are the ones considered here, is given as (see~\ref{app-1.2} for the proof) 
\begin{equation}
\label{ztilde}
\widetilde Z_{Neel}=B\Big(\frac{L}{2},\frac{L}{4}\Big).
\end{equation}
An overview of generic features of the overlaps is given in Figure~\ref{fig0:Neel-ov} 
(a) that shows  the squared N\'eel overlaps $|\langle\lambda|N\rangle|^2$ with the $XXX$ 
chain eigenstates $|\lambda\rangle$ versus the energy density $E/L$. The figure shows 
results for chains with $26\le L\le 38$ sites. The data are obtained by generating all 
the relevant parity-invariant BGT quantum numbers, and solving the associated BGT 
equations~\eref{bgt-eq}, to obtain the rapidities of $XXX$ chain eigenstates. Finally, 
the overlaps are calculated numerically using~\eref{Neel-ov}. Note that for $L=38$ 
from~\eref{ztilde} the total number of overlap shown in the Figure is $\widetilde Z_{Neel}
\sim 10^5$. 

Clearly, from Figure~\ref{fig0:Neel-ov} one has that the overlaps decay exponentially 
as a function of $L$, as expected. Moreover, at each finite $L$ a rapid decay as a 
function of $E/L$ is observed. The inset of Figure~\ref{fig0:Neel-ov} (a) (note the 
logarithmic scale on the $y$-axis) suggests that this decay is exponential. Complementary 
information is shown in Figure~\ref{fig0:Neel-ov} (b) reporting the histograms of $\kappa
\equiv-2\ln|\langle\lambda|N\rangle|/L$ (overlap distribution function). Larger values 
of $\kappa$, correspond to a faster decay with $L$ of the overlaps. The factor $1/L$ 
in the definition takes into account that the N\'eel overlaps typically vanish 
exponentially as $|\langle\lambda|N\rangle|^2\propto e^{-\kappa L}$ in the thermodynamic 
limit. Note that $\kappa$ is the driving term in the quench action approach (cf~\eref{Neel-ov-th}). 
As expected, from Figure~\ref{fig0:Neel-ov} (b) one has that the majority of the 
eigenstates exhibit small N\'eel overlap (note the maximum at $\kappa\sim 0.5$). 
Interestingly, the data suggest that $0.18\lesssim\kappa\lesssim0.7$. The vertical 
dash-dotted line in the figure is the $\kappa$ obtained from the N\'eel overlap of 
the ground state of the $XXX$ chain in the thermodynamic limit. This is derived using 
the ground state root distribution $\rho_1(\lambda)\propto1/\cosh(\pi\lambda)$~\cite{taka-book} 
and~\eref{Neel-ov-th}. On the other hand, the vertical dashed line denotes the N\'eel 
overlap $\sim 2/B(L,L/2)$ of the $S_z=0$ component of the ferromagnetic multiplet, which 
is at the top of the $XXX$ chain energy spectrum. Finally, the vertical dotted line in 
Figure~\ref{fig0:Neel-ov} (b) shows the quench action result for $\kappa$ in the thermodynamic 
limit. This is obtained by using~\eref{Neel-ov-th} and the saddle point root distributions 
$\rho_n^*$ (cf.~\eref{rho1-sp}-\eref{rho3-sp} for the results up to $n=3$). Note that $\kappa$ 
does not coincide with the peak of the overlap distribution function, as expected. This is  
due to the competition between the driving term~\eref{Neel-ov-th} and the Yang-Yang entropy 
$S_{YY}$ (cf.~\eref{qa-term}) in the quench action treatment of the N\'eel quench. 

%%%%%%%%%%%%%%%%%%%%%%%%%%%%%%%%%%%%%%%%%%%%%%%%%%%%%%%%%%%%%%%%%%%%%%%%%%%
\subsection{Overlap sum rules}
\label{sec:5.2}

%##################################################################
\begin{figure}[t]
\begin{center}
\includegraphics[width=.9\textwidth]{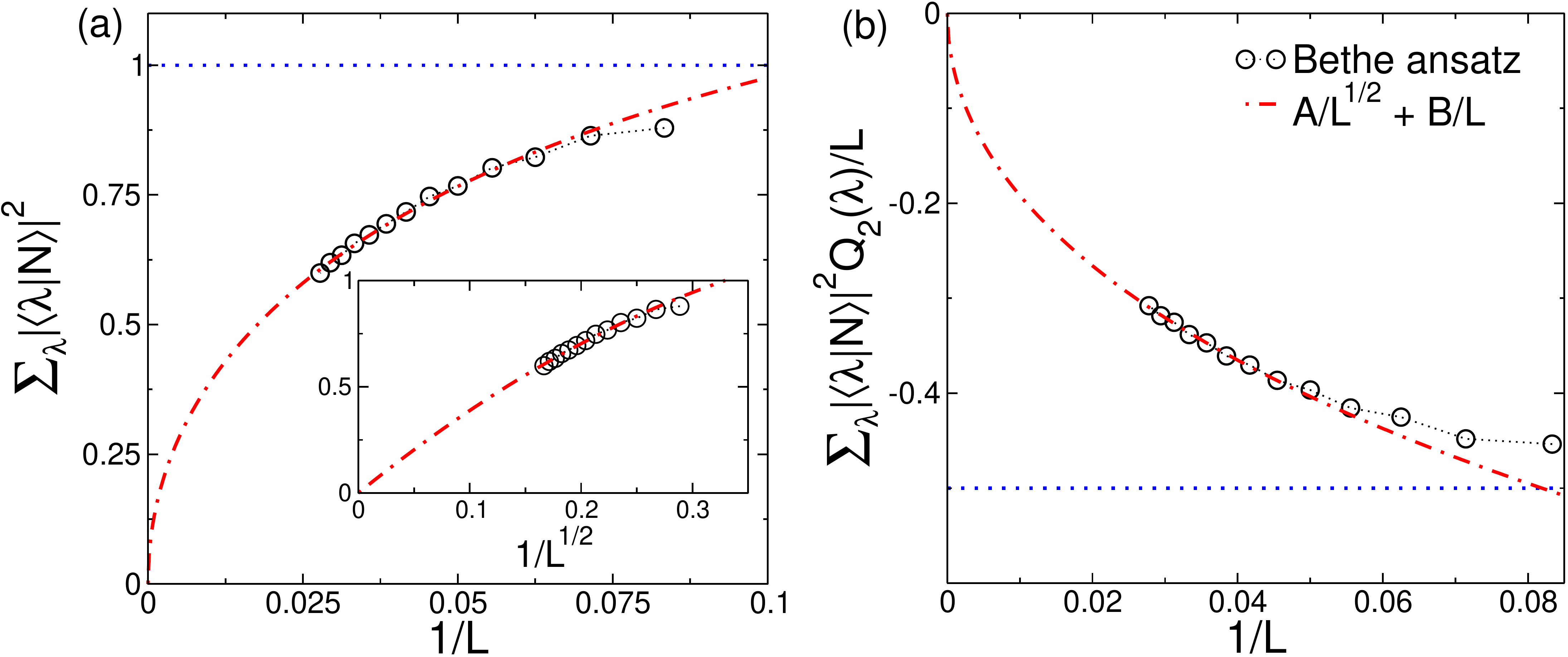}
\end{center}
\caption{Overlap sum rules for the N\'eel state $|N\rangle$: The role of 
 the zero-momentum strings. (a) The overlap sum rule $\sum_{\lambda}|
 \langle\lambda|N\rangle|^2=1$. Here $|\lambda\rangle$ are the eigenstates  
 of the $XXX$ chain. The $x$-axis shows the inverse chain length $1/L$. 
 The circles are Bethe ansatz results for chains up to $L=38$. 
 The data are obtained by  a full scanning of the  Hilbert space excluding eigenstates  
 with  zero-momentum strings. 
 The dotted line is the expected result at any $L$. 
 The data are compatible with a vanishing behavior in the thermodynamic limit. 
 The dash-dotted line is a fit to $A/L^{1/2}+B/L$, with $A,B$ fitting parameters. 
 Inset: The same data as in the main Figure now plotted versus $1/L^{1/2}$. 
 (b) The same as in (a) for the energy sum rule $\sum_{\lambda}|\langle
 \lambda|N\rangle|^2Q_2(\lambda)=Q_2^{(0)}$, with $Q_2(\lambda)$ the 
 energy of the eigenstate $|\lambda\rangle$ and $Q_2^{(0)}/L=-1/2$ the 
 N\'eel state energy density (dotted line in the Figure). 
}
\label{fig1:Neel-sr}
\end{figure}
%##################################################################

Here we study the effect of the zero-momentum strings eigenstates on the 
N\'eel overlap sum rules. We focus on the ``trivial'' sum rule, i.e., the 
normalization of the N\'eel state  
\begin{equation}
\label{sr-trivial}
\langle N|N\rangle=\sum\limits_{\lambda}|\langle\lambda|N\rangle|^2=1. 
\end{equation}
We also consider the N\'eel expectation value of the local conserved charge 
$Q_n$ of the $XXX$ chain (see subsection~\ref{sec:1.5}). These provide the 
additional sum rules
\begin{equation}
\label{sr-charge}
Q_n^{(0)}=\langle N|Q_n|N\rangle=\sum\limits_{\lambda}|\langle\lambda|N\rangle|^2
Q_{n}(\lambda)\quad\textrm{with}\quad n\in\mathbb{N}, 
\end{equation}
where $Q_n(\lambda)$ are the charges eigenvalues over the generic Bethe state 
$|\lambda\rangle$ (cf.~\eref{qngnk} and~\eref{gnk}). 
In~\eref{sr-charge} $Q_n^{(0)}$ is the expectation value of $Q_n$ over the 
initial N\'eel state. $Q_n^{(0)}$ have been calculated in Ref.~\cite{fagotti-2013} 
for any $n$. Due to the locality of $Q_n$, the translational invariance of the initial 
state, and the periodic boundary conditions, the density $Q_n^{(0)}/L$ does 
not depend on the length of the chain. 

Now we consider the sums \eref{sr-trivial} and~\eref{sr-charge} (for $n=2$, i.e., the energy sum rule)
restricted to the eigenstates with no zero-momentum strings.
These are shown in Figure~\ref{fig1:Neel-sr} (a) and (b), respectively. 
Note that $Q^{(0)}_2/L=-1/2$ in~\eref{sr-charge} (horizontal dotted line). 
The circles in Figure~\ref{fig1:Neel-sr} (a) are the Bethe ansatz results excluding 
the zero momentum strings. The data are the same as in Figure~\ref{fig0:Neel-ov}. 
The sum rules are plotted against the inverse chain length $1/L$, for $L\le 38$. 
As expected, both the sum rules are strongly violated, due to the exclusion of the to zero-momentum 
strings. Moreover, in both Figure~\ref{fig1:Neel-sr} (a) and (b) the data suggest a 
vanishing behavior upon increasing $L$. The dash-dotted lines are fits to  $A/L^{1/2}
+B/L$, with $A,B$ fitting parameters. Interestingly, the behavior as $\propto 
L^{-1/2}$ of the sum rules reflects that of the fraction of non-zero momentum string 
eigenstates $\widetilde Z_{Neel}/Z_{Neel}$. Specifically, from~\eref{zNeel1} and~\eref{ztilde} 
it is straightforward to derive that for $L\to\infty$
\begin{equation}
\label{beh}
\frac{\widetilde Z_{Neel}}{Z_{Neel}}\propto\frac{4}{\sqrt{\pi L}}. 
\end{equation}
% 

%##################################################################
\begin{figure}[t]
\begin{center}
\includegraphics[width=.9\textwidth]{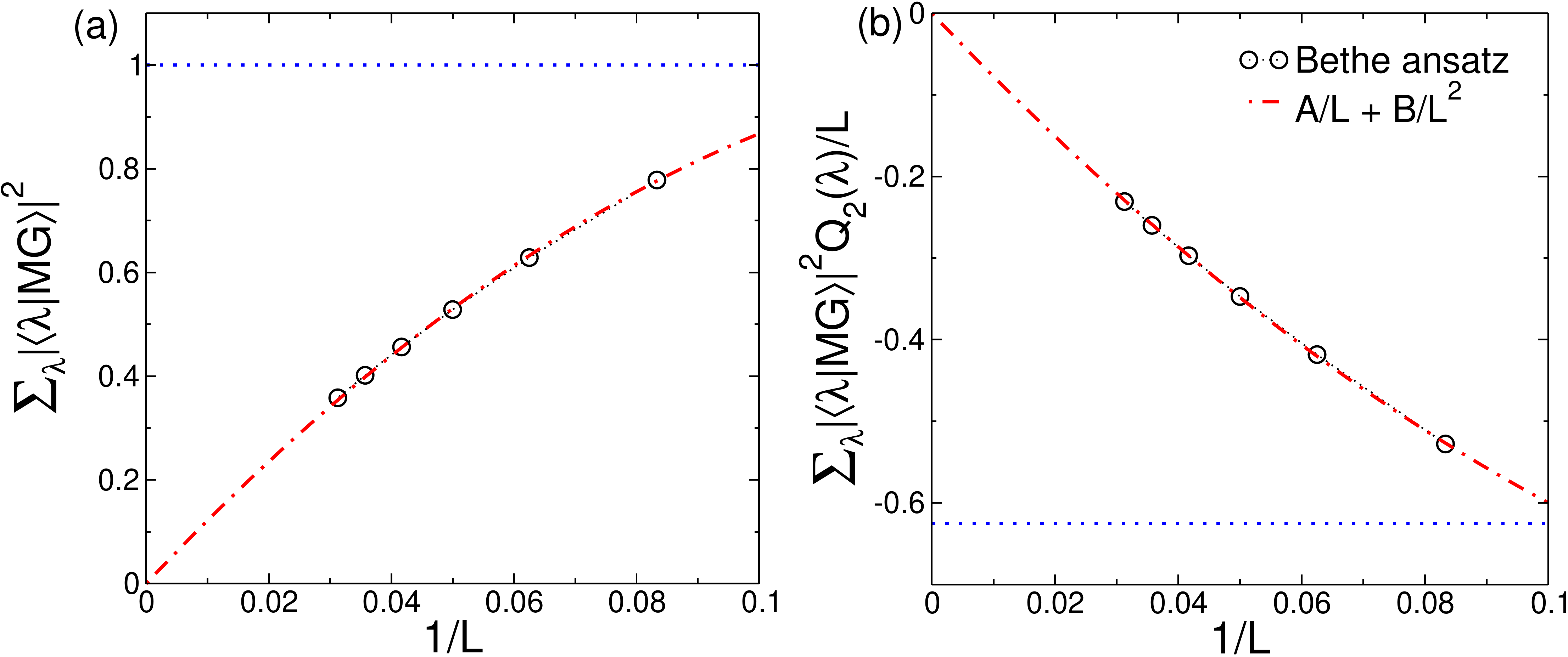}
\end{center}
\caption{Overlap sum rules for the Majumdar-Ghosh state $|MG\rangle$: 
 The role of the zero-momentum strings. (a) The sum rule $\sum_{\lambda}|
 \langle\lambda|MG\rangle|^2=1$, with $|\lambda\rangle$ the eigenstates   
 of the $XXX$ chain. The $x$-axis plots the inverse chain length $1/L$. 
 The circles are Bethe ansatz results for chains up to $L=32$. The results 
 are obtained by a full scanning of the chain Hilbert space excluding eigenstates with zero-momentum strings. 
 The dash-dotted line is a fit to $A/L+B/L^2$, with $A,B$ fitting parameters. 
 (b) The same as in (a) for the energy sum rule $\sum_{\lambda}|\langle\lambda|MG\rangle
 |^2Q_2(\lambda)=Q_2^{(0)}$, with $Q_2(\lambda)$  the energy of $|\lambda
 \rangle$ and $Q_2^{(0)}/L=-5/8$ the Majumdar-Ghosh energy density (dotted 
 line in the Figure). 
}
\label{fig2:dimer-sr}
\end{figure}
%##################################################################

The large $L$ behavior as $L^{-1/2}$  of the restricted sum 
rules is not generic, meaning that it depends on the pre-quench initial state $|\Psi_0
\rangle$. This is illustrated in Figure~\ref{fig2:dimer-sr}, focusing on the 
Majumdar-Ghosh (MG) state. As for the N\'eel state, only parity-invariant eigenstates 
can have non-zero Majumdar-Ghosh overlap. Their total number $Z_{MG}$ (cf.~\eref{p-inv-mg}) 
is given as 
\begin{equation}
\label{mg1}
Z_{MG}=B\Big(\frac{L}{2}-1,\frac{L}{4}-1\Big)+B\Big(\frac{L}{2}-1,
\frac{L}{4}-1\Big). 
\end{equation}
As for~\eref{zNeel1}, $Z_{MG}$ is only un upper bound for the number of Bethe states with 
non-zero Majumdar-Ghosh overlaps. Note also that at any size $L$ one has $Z_{MG}<Z_{Neel}$. 
This is due to the Majumdar-Ghosh state being invariant under $SU(2)$ rotations, since it 
contains only spin singlets. In contrast with the N\'eel state, this implies that the 
Majumdar-Ghosh stat has non-zero overlap only with the $S^z_T=0$ sector of the $XXX$ chain 
spectrum. After restricting to the situation with no zero-momentum strings, the total number 
of parity-invariant eigenstates $\widetilde Z_{MG}$ in the sector with $S_T^z=0$ is now 
(cf.~\eref{mg-fi})
\begin{equation}
\label{mg2}
\widetilde Z_{MG}=B\Big(\frac{L}{2},\frac{L}{4}\Big)-B\Big(\frac{L}{2},
\frac{L}{4}-1\Big). 
\end{equation}
Panels (a) and (b) in Figure~\ref{fig2:dimer-sr} report the restricted sums~\eref{sr-trivial} 
and~\eref{sr-charge} for the Majumdar-Ghosh state. The data are obtained using the 
analytic results for the overlaps in subsection~\ref{sec:2.3}. The expected value for 
the energy density sum rule is $Q_2^{(0)}=-5/8$ (horizontal dotted line in 
Figure~\ref{fig2:dimer-sr} (b)). Similar to Figure~\ref{fig1:Neel-sr}, due to the exclusion 
of the zero-momentum strings, the sum rules are violated, exhibiting vanishing behavior in 
the thermodynamic limit. However, in contrast with the N\'eel case, one has the behavior 
as $1/L$, as confirmed by the fits (dash-dotted lines in Figure~\ref{fig2:dimer-sr}). 
The vanishing of the sum rules in the thermodynamic limit 
reflects the behavior of $\widetilde Z_{MG}/Z_{MG}$ as (see~\eref{mg1} and~\eref{mg2})
\begin{equation}
\frac{\widetilde Z_{MG}}{Z_{MG}}=\frac{4}{4+L}. 
\end{equation}

\subsection{Quench action reweighting}

The results in the previous section could lead to the {\it erroneous} conclusion that the states with zero momentum strings 
are essential in order to reconstruct the thermodynamic values of any observables, in stark contrast with the quench action 
results \cite{pozsgay-2014A,wouters-2014A} that fit perfectly with the numerical simulations reported in the same papers. 
The solution to this apparent paradox is understood within the quench action formalism, 
which predicts that the {\it representative state} of the stationary state in the thermodynamic limit can be completely described 
restricting to the states without zero momentum strings. 
However, when reconstructing this state (both analytically and numerically), it must be properly normalised to unity in the 
reduced subspace we are considering. 
Consequently, we expect that after a quench from the initial state $\Psi_0$ the stationary expectation value 
of any local observables ${\cal O}$ can rewritten as 
\begin{equation}
\langle{\cal O}\rangle= \frac{\sum_{\lambda} |\langle \lambda| \Psi_0\rangle|^2   \langle\lambda|{\mathcal O}|\lambda\rangle }{ 
\sum_{\lambda} |\langle \lambda| \Psi_0\rangle|^2}, 
\qquad w_{\Psi_0}\equiv \sum_{\lambda} |\langle \lambda| \Psi_0\rangle|^2\,,
\label{rewe}
\end{equation}
where, crucially, the sums over $\lambda$ are restricted to the states without zero momentum strings. 
This means that any local observable must be reweighed by the factor $w_{\Psi_0}$, reflecting the fact that we are considering 
only a small portion of the total Hilbert space.

\begin{figure}[t]
\begin{center}
\includegraphics[width=.95\textwidth]{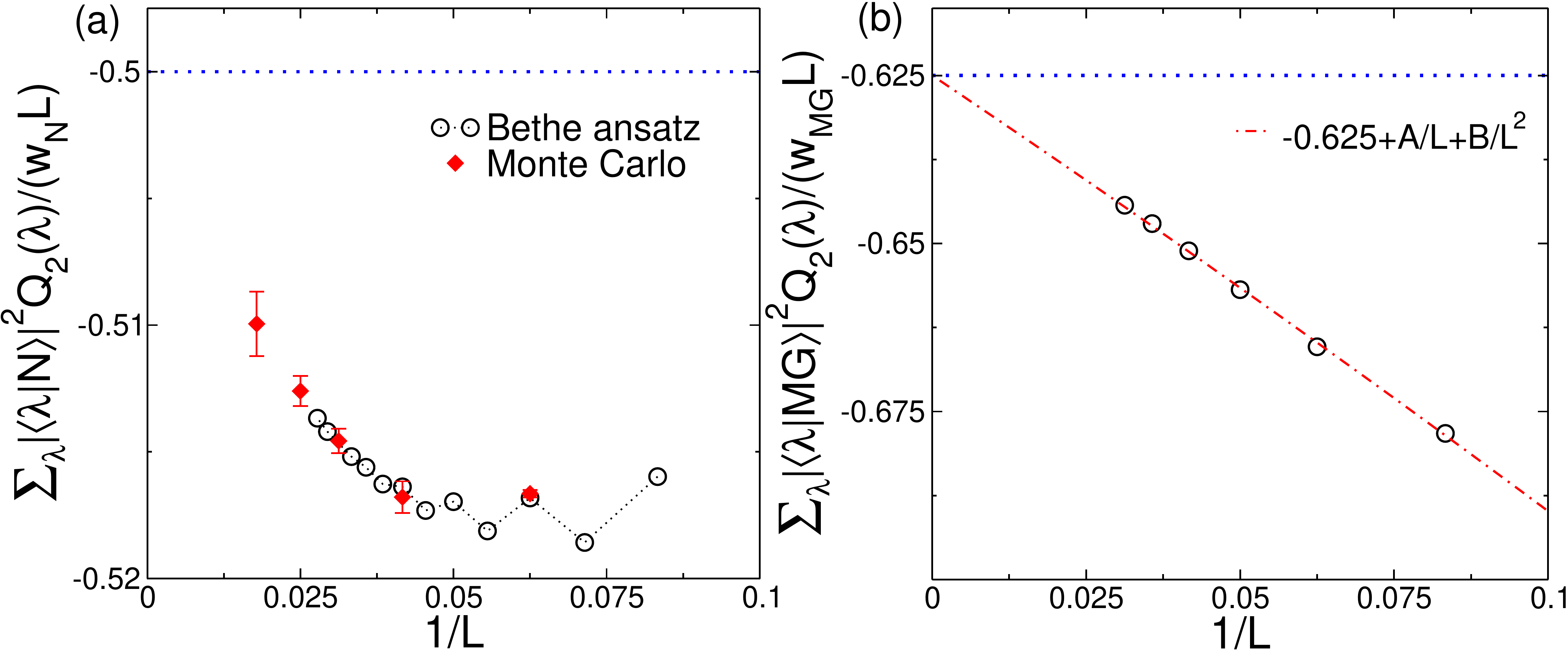}
\end{center}
\caption{Energy density expectation values for a quench from N\'eel (a) and Majumdar-Ghosh state (b) obtained by means of the 
reweighting \eref{rewe}.
For the Majumdar-Ghosh state a quadratic fit (forcing the asymptotic value $Q_2^{(0)}=-5/8$) describes perfectly the reweighed 
data.
For the N\'eel state we also report the Monte Carlo data obtained in the next section, showing perfect agreement with the 
reweighed Bethe ansatz data. 
}
\label{fig:rew}
\end{figure}

The results obtained by reweighting the data in Figs. \ref{fig1:Neel-sr} and \ref{fig2:dimer-sr}
are shown in Fig. \ref{fig:rew} focusing on the conserved energy density for both the N\'eel and 
Majumdar-Ghosh initial states. 
It is evident that the data after reweighting approach a finite value for large $L$ in contrast with the bare vanishing values 
(cf. Figs.  \ref{fig1:Neel-sr} and \ref{fig2:dimer-sr}). 
It is, however, also clear that this procedure introduces some finite size corrections due to the restriction of the Hilbert space that is 
only justified in the thermodynamic limit. 
For the Majumdar-Ghosh state these corrections are monotonous and one can extrapolate with a quadratic fit in $1/L$
to infinite size, in order to reproduce the correct thermodynamic expectation (see Fig. \ref{fig:rew} panel (b)).
For the N\'eel initial state instead, the corrections to the scaling are not monotonous and an extrapolation 
based only on the data up to $L=38$ are inconclusive. 
In Fig. \ref{fig:rew} we plot on top of the exact Bethe ansatz data, also Monte Carlo results obtained by sampling the restricted 
Hilbert space (see next section). 
The Monte Carlo allows to consider larger system sizes, which turns out to be enough for a good extrapolation, as we
will discuss in the next section. 
We limit here to notice that the agreement between exact and Monte Carlo data is excellent, corroborating 
the correctness of both approaches.  

%%%%%%%%%%%%%%%%%%%%%%%%%%%%%%%%%%%%%%%%%%%%%%%%%%%%%%%%%%%%%%%%%%%%%%%%%%%
\section{Monte Carlo implementation of the quench action approach}
\label{sec6:mcqa}

In this section, by generalizing the results in~\cite{alba-2015}, we present 
a Monte Carlo implementation of the quench action approach for the N\'eel quench in 
the $XXX$ chain. The key idea is to sample the eigenstates of the finite-size $XXX$ 
chain with the quench action probability distribution, given in~\eref{qa-d-ensemble}. 
Similar Monte Carlo techniques to sample the Hilbert space of integrable models 
have been used in Ref.~\cite{gu-2005,buccheri-2011,faribault-2013}. 
Importantly, here we consider a truncated Hilbert space, restricting ourselves to the 
eigenstates corresponding to solutions of the BGT equations with no zero-momentum 
strings. Our main physical result is that, despite this restriction, the remaining 
eigenstates contain enough information to correctly reproduce the post-quench 
thermodynamic behavior of the $XXX$ chain. 

In subsection~\ref{sec:6.1} we detail the Monte Carlo algorithm. In 
subsection~\ref{sec:6.2} we numerically demonstrate that after the Monte Carlo 
``resampling'' the N\'eel sum rules~\eref{sr-charge} are restored, in the 
thermodynamic limit. The Hilbert space truncation is reflected only in 
$\propto 1/L$ finite-size corrections to the sum rules. In the Bethe ansatz 
language the eigenstates sampled by the Monte Carlo become equivalent to the 
quench action representative state in the thermodynamic limit. Here this is 
explicitly demonstrated by numerically extracting the quench action root 
distributions $\pmb{\rho}^*$ (cf.~\eref{rho1-sp}-\eref{rho3-sp}). The 
numerical results are found in remarkable agreement with the quench action.

%%%%%%%%%%%%%%%%%%%%%%%%%%%%%%%%%%%%%%%%%%%%%%%%%%%%%%%%%%%%%%%%%%%%%%%%%%%
\subsection{The quench action Monte Carlo algorithm}
\label{sec:6.1}

The Monte Carlo procedure starts with a randomly selected parity-invariant eigenstate 
(Bethe state) of the $XXX$ chain, in the sector with zero magnetization, i.e., $M=L/2$ 
particles. As the N\'eel state is not invariant under $SU(2)$ rotations, in order to 
characterize the Bethe states one has to specify the number $N_{\infty}$ of infinite 
rapidities (see~\ref{sec:1.2}). The number of remaining particles corresponding to finite 
BGT rapidities $M'$ is $M'=L/2-N_\infty$. The Bethe state is identified by a parity-invariant 
BGT quantum number configuration that we denote as ${\mathcal C}$. Due to the parity-invariance 
and the zero-momentum strings being excluded, ${\mathcal C}$ is identified by the 
number $m'$ of parity-invariant quantum numbers $\{\pm I_j\}_{j=1}^{m'}$ (equivalently, 
root pairs $\{\pm\lambda_j\}_{j=1}^{m'}$). The string 
content associated with the state is denoted as $\widetilde{\mathcal S}=\{\tilde s_1,
\dots,\tilde s_{m'}\}$, where $\tilde s_n$ is the number of pairs of $n$-strings. The 
Monte Carlo procedure generates a new parity-invariant eigenstate of the $XXX$ chain, 
and it consists of four steps: 
\begin{enumerate}
\item[\circled{1}] Choose a new number of finite-momentum particles $M''$ and of 
parity-invariant rapidity pairs $m''\equiv M''/2$ with 
probability ${\mathcal P}(M'')$ as 
\begin{equation}
\label{PM}
{\mathcal P}(M'')=\frac{\widetilde Z'_{Neel}(L,M'')}
{\widetilde{Z}_{Neel}(L)}, 
\end{equation}
where $\widetilde Z_{Neel}(L)$ is defined in~\eref{ztilde}, and $\widetilde 
Z'_{Neel}$ is the number of parity-invariant eigenstates with no zero-momentum 
strings in the sector with fixed particle number $M''$  (cf.~\eref{Neel-fi} for 
the precise expression). 
\item[\circled{2}] Choose a new string content $\widetilde{\mathcal S}'\equiv
\{\tilde s_1',\dots,\tilde s'_{m''}\}$ with probability ${\mathcal P}'(M'',
\widetilde{\mathcal S}')$
\begin{equation}
\label{PS}
{\mathcal P}'(M'',\widetilde{\mathcal S}')=\frac{1}{\widetilde Z'_{Neel}
(L,M'')}\prod_{n=1}^{m''}B\Big(\frac{L}{2}-\sum\limits_{l=1}^{
m''}t_{nl}\tilde s'_l,\tilde s'_n\Big), 
\end{equation}
where the matrix $t_{nl}$ is defined in~\eref{bt-qn-bound}.
\item[\circled{3}] Generate a new parity-invariant quantum number configuration 
${\mathcal C}'$ compatible with the $\widetilde {\mathcal  S}'$ obtained in step 
$\circled{2}$. Solve the corresponding BGT equations~\eref{bgt-eq}, finding the 
rapidities $\{\pm\lambda'_j\}_{j=1}^{m''}$ of the new parity-invariant eigenstate. 
\item[\circled{4}] Calculate the N\'eel overlap $\langle\{\pm\lambda'_j\}_{j=1}^{m''}
|N\rangle$ for the new eigenstate,  using~\eref{Neel-ov}~\eref{red-G+}~\eref{red-G-} 
and~\eref{Neel-k}. Accept the new eigenstate with the quench action Metropolis 
probability 
\begin{equation}
\label{metropolis}
{\mathcal P}''_{\lambda\to\lambda'}=\textrm{Min}\Big\{1,\exp\Big(-
2{\rm Re}({\mathcal E}'-{\mathcal E})\Big)\Big\}, 
\end{equation}
where ${\mathcal E}'\equiv-\ln\langle\{\pm\lambda'_j\}_{j=1}^{m''}
|N\rangle$, ${\mathcal E}\equiv-\ln\langle\{\pm\lambda_j\}_{j=1}^{m''}
|N\rangle$. 
\end{enumerate}
Note that while the steps $1$-$3$ account for the string content and particle 
number probabilities of the parity-invariant states, step $4$ assigns to the 
different eigenstates the correct quench action probability. 

For a generic local observable ${\mathcal O}$, its quench action expectation $\langle{\mathcal O}
\rangle$ is obtained as the arithmetic average 
of the eigenstates expectation values $\langle\lambda|{\mathcal O}|\lambda\rangle$, 
with $|\lambda\rangle$ the eigenstates sampled by the Monte Carlo, as 
\begin{equation}
\label{qamc-obs}
\langle{\mathcal O}\rangle=\frac{1}{N_{mcs}}\sum\limits_{\lambda}\langle\lambda|
{\mathcal O}|\lambda\rangle. 
\end{equation}
Here $N_{mcs}$ is the total number of Monte Carlo steps. Note that, as usual in 
Monte Carlo, some initial steps have to be neglected to ensure equilibration. 
Note that~\eref{qamc-obs} can be used for any observable ${\mathcal O}$ for 
which the the Bethe state expectation value $\langle\lambda|{\mathcal O}|\lambda
\rangle$ (form factor) is known. 

Finally, it is worth stressing that although the Monte Carlo sampling is done only on the 
zero-momentum free Hilbert subspace, the algorithm does not suffer of the reweighting problems 
found in the exact method (see the preceding section), because the expectation values 
\eref{qamc-obs} are automatically normalised by the factor $N_{mcs}$.

%%%%%%%%%%%%%%%%%%%%%%%%%%%%%%%%%%%%%%%%%%%%%%%%%%%%%%%%%%%%%%%%%%%%%%%%%%%
\subsection{The N\'eel overlap sum rules: Monte Carlo results}
\label{sec:6.2}

%##################################################################
\begin{figure}[t]
\begin{center}
\includegraphics[width=.9\textwidth]{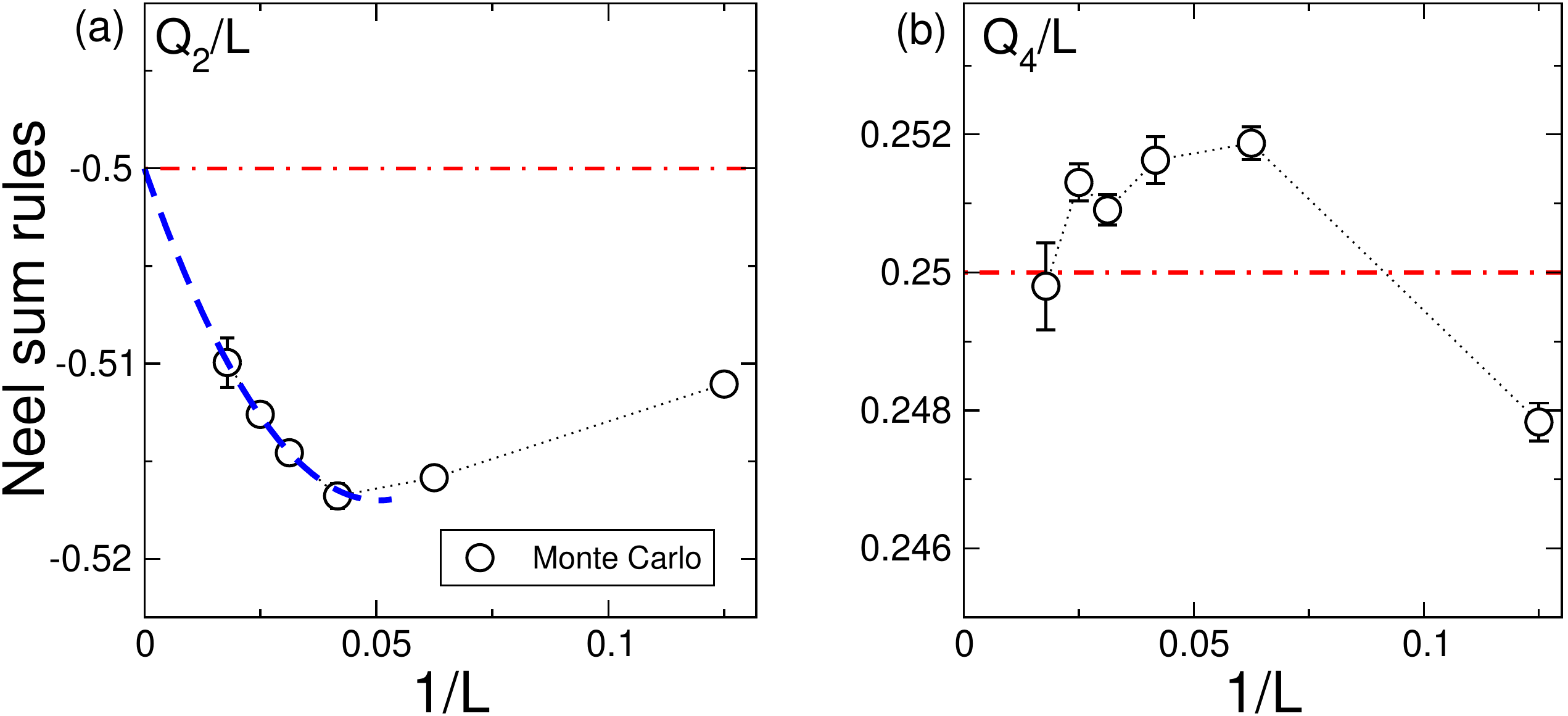}
\end{center}
\caption{The overlap sum rules for the N\'eel state $|N\rangle$ in the 
 Heisenberg spin chain: Numerical results obtained by the Monte Carlo 
 sampling of the chain Hilbert space. In all panels the $x$ axis shows 
 the inverse chain length $1/L$ (a) The energy sum rule $\sum_\lambda|
 \langle N|\lambda\rangle|^2Q_2(\lambda)/L=Q^{(0)}_2$, with $|\lambda
 \rangle$ the generic eigenstate of the $XXX$ chain, $Q_2(\lambda)/L$ 
 the associated energy density, and $Q_2^{(0)}=-1/2$ the N\'eel energy 
 density. The symbols are Monte Carlo data. The dash-dotted line is the 
 expected result $Q_2^{(0)}$. The dashed line is a fit to the behavior 
 $-1/2+A/L+B/L^2$, with $A,B$ fitting parameters. (b) Same as in (a) for 
 the charge $Q_4$ without the fit. 
}
\label{fig3:Neel-qamc-sr}
\end{figure}
%##################################################################

The validity of the Monte Carlo approach outlined in~\ref{sec:6.2} is demonstrated 
in Fig.~\ref{fig3:Neel-qamc-sr}. The Figure focuses on the N\'eel overlap 
sum rules for the conserved charges densities $Q_2/L$ and $Q_4/L$ (cf. 
subsection~\ref{sec:1.5} for the definition of the charges, and~\eref{sr-charge} 
for the associated sum rules). 
Note that in~\eref{sr-charge}  the sum is now over the 
eigenstates $|\lambda\rangle$ sampled by the Monte Carlo. 
Panel (a) in Figure~\ref{fig3:Neel-qamc-sr} reports the sum rule for 
the energy density $Q_2/L$ (these are the same data reported in Fig. \ref{fig:rew} which 
perfectly agree with the exact data, corroborating the accuracy of the Monte Carlo sampling). 
The circles in the Figure are Monte Carlo data for the Heisenberg chain with $L\le 56$ 
sites. The data correspond to Monte Carlo simulations with $N_{mcs}\sim 10^7$ Monte 
Carlo steps (mcs). In all panels the $x$-axis shows the inverse chain length $1/L$. 

Clearly, the Monte Carlo data suggest that in the thermodynamic limit the N\'eel 
overlap sum rules~\eref{sr-charge} are restored, while violations are present for 
finite chains. This numerically confirms that the truncation of the Hilbert space, 
i.e., removing the zero-momentum strings, gives rise only to scaling corrections, 
while the thermodynamic behavior after the quench is correctly reproduced. 
Note that the data in panel (a) are suggestive  of the behavior $\propto 1/L$ for 
the scaling corrections, as confirmed by the fit to $-1/2+A/L+b/L^2$ (dashed line in the Figure), 
with $A,B$ fitting parameters.

Similarly, panel (b) in Figure~\ref{fig3:Neel-qamc-sr} reports the charge 
density $Q_4/L$. Also in this case, the Monte 
Carlo data for $L=48$ are already compatible with the expected result $Q_4/L=1/4$ 
in the thermodynamic limit. 
The scaling corrections  are however not monotonous and it is impossible to 
proceed to a proper extrapolation to the thermodynamic limit. 
This could be attributed to fact that the support of $Q_n$, i.e., the number of 
sites where the operator acts non trivially, increases linearly with $n$ 
(see~\cite{grabowski-1995} for the precise expression). 
Anyhow, notice that all data deviate from the expected asymptotic value for about 1\%. 

%%%%%%%%%%%%%%%%%%%%%%%%%%%%%%%%%%%%%%%%%%%%%%%%%%%%%%%%%%%%%%%%%%%%%%%%%%%
\subsection{Extracting the quench action root distributions}
\label{sec:6.3}

%##################################################################
\begin{figure}[t]
\begin{center}
\includegraphics[width=.95\textwidth]{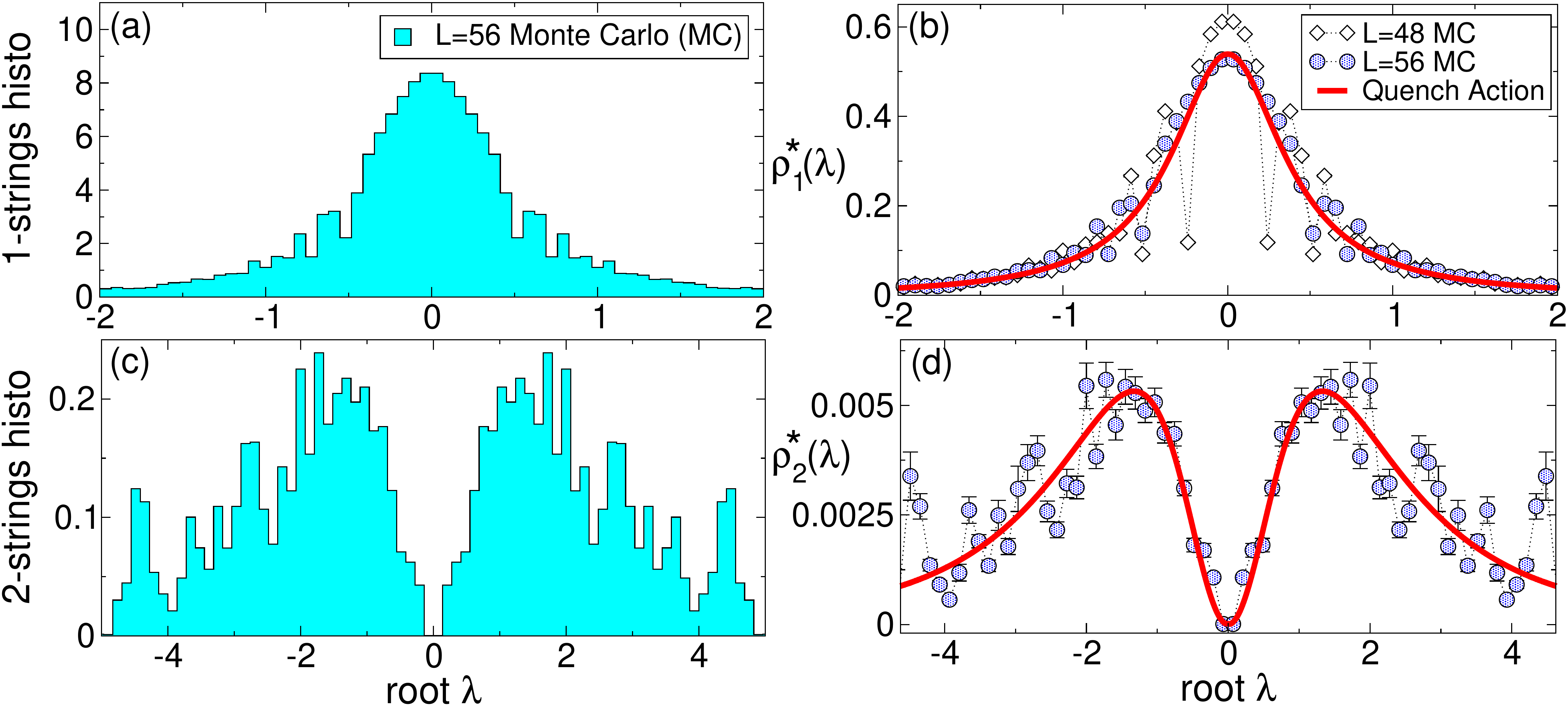}
\end{center}
\caption{ The quench action root distributions $\rho^*_1(\lambda)$ and 
 $\rho^*_2(\lambda)$ for the $1$-strings and $2$-strings, respectively: Monte 
 Carlos results. (a) The histograms of the $1$-string Bethe-Gaudin-Takahashi 
 (BGT) roots $\lambda$ sampled in the Monte Carlo. The data are for a chain 
 with $L=56$ sites and a Monte Carlo history with $N_{mcs}\sim 10^7$ Monte 
 Carlo steps. The $y$-axis is divided by a factor $10^6$ for convenience. 
 %The width of the histogram bin is $\Delta\lambda\sim 0.07$. 
 (c) The same as in (a) for the $2$-string roots. (b) The $1$-string root distribution 
 $\rho^*_1(\lambda)$ plotted versus $\lambda$ for two chains with $L=48$ 
 and $L=56$ (diamond and circles, respectively). The full line 
 is the quench action analytic result in the thermodynamic limit. (d) 
 The same as in (b) for the $2$-string root distribution $\rho^*_2(\lambda)$. 
 In both (b) and (d) the oscillations are finite-size effects, whereas 
 the error bars are the statistical Monte Carlo errors. 
}
\label{fig4:Neel-rho}
\end{figure}
%##################################################################

The BGT root distributions corresponding to the quench action steady state 
(cf.~\eref{rho1-sp}-\eref{rho3-sp}) $\pmb{\rho^*}=\{\rho^*_n(\lambda)\}_{n=
1}^{\infty}$ can be extracted from the Monte Carlo simulation,  similar to 
what has been done in Ref.~\cite{alba-2015} for the Generalized Gibbs Ensemble 
(GGE) representative state. The idea is that for the local observables considered 
here, in each eigenstate expectation value $\langle\lambda|{\mathcal O}|\lambda
\rangle$ in~\eref{qamc-obs} one can isolate the contribution of the different 
string sectors as 
\begin{equation}
\label{ref-form}
\langle\lambda|{\mathcal O}|\lambda\rangle=\sum\limits_{n,\gamma}{\mathcal O}_n
(\lambda_{n;\gamma}). 
\end{equation}
Here ${\mathcal O}_n$ is the contribution of the BGT $n$-strings to the expectation 
value of ${\mathcal O}$, and $\pm\lambda_{n;\gamma}$, with $\gamma$ labeling the 
different $n$-strings, are the solutions of the BGT equations~\eref{bgt-eq} 
identifying the Bethe state $|\lambda\rangle$. We should stress again that~\eref{ref-form} is true only 
for local or quasi-local observables, while generic observables are more complicated 
functions of the rapidities. By comparing~\eref{qamc-obs} 
and~\eref{obs-th} one obtains that in the limit $L,N_{mcs}\to\infty$ 
\begin{equation}
\quad\fl\lim_{N_{mcs}\to\infty}\,\frac{1}{N_{mcs}}\sum\limits_{\lambda_{n;\gamma}}
{\mathcal O}_n(\lambda_{n;\gamma})\,\stackrel{L\to\infty}{\overrightarrow{\hspace{40pt}}}
\,\langle\pmb{\rho^*}|{\mathcal O}|\pmb{\rho^*}\rangle\equiv\sum_n\int_{-\infty}^{+\infty}d
\lambda\rho^*_{n}(\lambda){\mathcal O}_n(\lambda). 
\end{equation}
This suggests that the histogram of the $n$-strings BGT roots sampled in the Monte 
Carlo converges in the thermodynamic limit to the saddle point root distribution 
$\rho^*_n(\lambda)$. 

This is demonstrated numerically in Figure~\ref{fig4:Neel-rho} considering $\rho^*_1(\lambda)$ 
(panels (a)(b)) and $\rho^*_2(\lambda)$ (panel (c)(d)). The histograms correspond to Monte 
Carlo data for $L=48$ and $L=56$ sites. Panel (a) and (c) show the  histograms of the 
$1$-string and $2$-string BGT roots sampled in the Monte Carlo. The $y$-axis is rescaled 
by a factor $10^6$ for convenience. The width of the histogram bins $\Delta\lambda$ is 
$\Delta\lambda\approx0.02$ and $\Delta\lambda\approx0.001$ for $\rho_1(\lambda)$ and 
$\rho_2(\lambda)$, respectively. The histogram fluctuations are due both to the finite 
statistics (finite $N_{mcs}$) and to the finite size of the chain. 

The extracted quench-action root distributions $\rho^*_1(\lambda)$ and $\rho^*(\lambda)$ 
are shown in panels (b) and (d). The data are the same as in panel (a)(c). The normalization 
of the distributions is chosen such as to match the analytical results from~\eref{rho1-sp} 
and~\eref{rho2-sp}, i.e., $\int d\lambda\rho^*_1(\lambda)\approx0.31$ and $\int d\lambda
\rho^*_2(\lambda)\approx0.015$. The Monte Carlo error bars shown in the Figure are obtained 
with a standard jackknife analysis~\cite{quenouille-1949,wolff-2004}. 
The continuous lines are the expected analytic results in the thermodynamic limit 
(cf.~\eref{rho1-sp}~\eref{rho2-sp}). 

Clearly, the Monte Carlo data are in excellent agreement with~\eref{rho1-sp} in the whole 
range $-2\le\lambda\le2$ considered. For $\rho^*_1(\lambda)$ the statistical error bars 
are smaller than the symbol size. The oscillating corrections around $|\lambda|\sim0.5$ 
are lattice effects, which decrease with increasing the chain size 
(see the data for $L=48$ in the Figure). Much larger finite-size effects are observed for 
$\rho^*_2(\lambda)$ (panel (d) in the Figure). Specifically, the corrections are larger on 
the tails of the root distribution. Moreover, the Monte Carlo error bars are clearly 
larger than for $\rho_1^*(\lambda)$. This is due to the fact that since $\int d\lambda
\rho^*_2(\lambda)/\sum_n\int d\lambda\rho_n^*(\lambda)\approx 0.04$, the Monte Carlo 
statistics available for estimating $\rho_2^*(\lambda)$ is effectively reduced as 
compared to $\rho_1^*(\lambda)$. Finally, we numerically observed that finite-size 
corrections and Monte Carlo error bars are even larger for the $3$-strings root 
distribution $\rho^*_3(\lambda)$, which makes its numerical determination more difficult.

%%%%%%%%%%%%%%%%%%%%%%%%%%%%%%%%%%%%%%%%%%%%%%%%%%%%%%%%%%%%%%%%%%%%%%%%%%%
\section{Conclusions}
\label{conclusions}

We developed a finite size implementation of the quench action method for integrable models. 
We focused on the spin-$1/2$ isotropic Heisenberg ($XXX$) chain, 
considering simple product states as initial states, but the approach is of general validity. 
The main ingredient of the approach is the knowledge of the overlaps between the pre-quench
states and the $XXX$ chain  eigenstates (that for the cases at hand have been obtained recently~\cite{
pozsgay-2014a,brockmann-2014,brockmann-2014b,brockmann-2014c,piroli-2014}). 
For chains up to about 40 spins the (relevant part of the) Hilbert space can be fully spanned, while 
for large systems we performed an effective Monte Carlo sampling. 
The main outcome of the method is a precise determination of the root distributions which allow
then the determination of local observables by standard methods. 
Thermodynamic quantities are finally extracted using finite-size scaling. 
The main result of this papers have been already summarised in the introductory section and 
we limit here to some discussions about further developments. 

First of all, 
the importance of the proposed method relies on the property that the only required ingredient is the (analytical or even numerical)
knowledge of the overlaps between the initial state and the Bethe states. 
For this reason, it can be implemented even in cases in which an analytic thermodynamic solution is not 
available. 

Another interesting consequence of our work is that we can use a vanishing fraction of the eigenstates in order 
to determine the thermodynamic behavior. It is clearly interesting to understand whether there are other clever ways 
to further reduce the fraction of considered states (without knowing the exact solution, when we can just 
pinpoint one representative eigenstate). 

Finally, it is an open interesting issue to understand how the present method can be used to describe the time evolution 
of a finite but large system and in particular how to reconstruct the time evolution from a vanishing fraction of relevant 
eigenstates, eventually mimicking  the strategy employed in thermodynamic limit \cite{de-nardis-2015a,de-nardis-2015}.

%%%%%%%%%%%%%%%%%%%%%%%%%%%%%%%%%%%%%%%%%%%%%%%%%%%%%%%%%%%%%%%%%%%%%%%%%%%
\section*{Acknowledgments}

We are very grateful to Fabian Essler for collaboration at the beginning of this project 
and for very fruitful discussions. 
We thank Maurizio Fagotti for useful discussions in the early stage of this manuscript 
and for providing us the exact diagonalization results reported in the appendices. 
We thank Lorenzo Piroli for useful discussions and comments. 
All authors acknowledge support by the ERC under Starting Grant 279391 EDEQS. 
%This work was done in part when FHL was visiting SISSA, whose hospitality is kindly acknowledged. 
%This work was supported by the EPSRC under grants EP/I032487/1 and EP/J014885/1 (FHLE).

%%%%%%%%%%%%%%%%%%%%%%%%%%%%%%%%%%%%%%%%%%%%%%%%%%%%%%%%%%%%%%%%%%%%%%%%%%%
\appendix

%%%%%%%%%%%%%%%%%%%%%%%%%%%%%%%%%%%%%%%%%%%%%%%%%%%%%%%%%%%%%%%%%%%%%%%%%%%
\section{Exact N\'eel and Majumdar-Ghosh overlaps for a small Heisenberg chain} 
\label{app-L12}

In this section we provide exact diagonalization results for the overlaps between the 
N\'eel state and the Majumdar-Ghosh (MG) state and all the eigenstates of the Heisenberg 
spin chain with $L=12$ sites. For the eigenstates without zero-momentum strings, we 
also provide the overlaps obtained using the string hypothesis~\eref{Neel-ov}\eref{mg-ov}. 
This allows to check the validity of the string hypothesis when calculating overlaps. 
Moreover, this also provides a simple check of the counting formula~\eref{N-count-nz}. 

%%%%%%%%%%%%%%%%%%%%%%%%%%%%%%%%%%%%%%%%%%%%%%%%%%%%%%%%%%%%%%%%%%%%%%%%%%%
\subsection{N\'eel overlap}
\label{app-Neel}

%%%%%%%%%%%%%%%%%%%%%%%%%%%%%%%%%%%%%%%%%%
\begin{table}[ht]
\scriptsize
\centering
Bethe states with nonzero N\'eel overlap ($L=12$)\\[1ex]
\begin{tabular}{rrrrrr}
\toprule
String content & $2I_n$ & $q$ & E & $|\langle\lambda|N\rangle|^2$ (exact) & $|\langle
\lambda|N\rangle|^2$ (BGT) \\[0.3em]
\toprule
6 inf & - & - & $0$ & $0.002164502165$ & $0.002164502165$\\
\midrule
\{2,0\}\, 4 inf &$1_1$ & $2$ & $-3.918985947229$ & $0.096183409244$ & $0.096183409244$\\
 &$3_1 $ & & $-3.309721467891$ & $0.011288497947$ &             $0.011288497947$\\
 &$5_1 $ & & $-2.284629676547$ & $0.004542580506$ &             $0.004542580506$\\
 &$7_1 $ & & $-1.169169973996$ & $0.002752622983$ &             $0.002752622983$\\
 &$9_1 $ & & $-0.317492934338$ & $0.002116006203$ &             $0.002116006203$\\
\midrule
\{4,0,0,0\}\, 2 inf &$1_1 3_1 $ & $4$ & $-7.070529325964$ & $0.310133033838$ &$0.310133033838$\\
  &$1_1 5_1 $ & & $-5.847128730477$ & $0.129277023687$ &           $0.129277023687$\\
  &$ 1_1 7_1$ & & $-4.570746557876$ & $0.085992436024$ &           $0.085992436024$\\
  &$ 3_1 5_1$ & & $-5.153853093221$ & $0.015256395523$ &           $0.015256395523$\\
  &$3_1 7_1 $ & & $-3.916336243695$ & $0.010091113504$ &           $0.010091113504$\\
  &$5_1 7_1 $ & & $-2.817696043731$ & $0.004059780228$ &           $0.004059780228$\\
\midrule
\{0,2,0,0\}\, 2 inf &$1_2 $ & $2$ & $-1.905667167442$ & $0.001207238321$ & $0.0012072{\color{red}45406}$\\
  &$3_2 $ & & $-1.368837200825$ & $0.002340453815$ &            $0.0023{\color{red}25724713}$\\
  &$5_2 $ & & $-0.681173793635$ & $0.001921010489$ &            $0.0019{\color{red}39001396}$\\
\midrule
\{1,0,1,0\}\, 2 inf &$0_1 0_3$ & $2$ & $-2.668031843135$ & $0.034959609810$ & -\\
\midrule
\{6,0,0,0,0,0\}\, 0 inf &$1_1 3_1 5_1$ & $6$ & $-8.387390917445$ & $0.153412152966$ & $0.153412152966$\\
\midrule
\{2,2,0,0,0,0\}\, 0 inf &$1_1 1_2$ & $4$ & $-5.401838225870$ & $0.040162686361$ & $0.04{\color{red}1042488913}$\\  
&$3_1 1_2 $ & & $-4.613929948329$ & $0.004636541934$ & $0.004{\color{red}730512604}$\\
  &$5_1 1_2 $ &  & $-3.147465758841$ & $0.001335522556$ & $0.00133{\color{red}7334035}$\\
\midrule
\{3,0,1,0,0,0\}\, 0 inf &$0_1 2_1 0_3$ & $4$ & $-6.340207488736$ & $0.052743525774$ & -\\
  &$0_1 4_1 0_3$ & & $-5.203653009936$ & $0.015022005621$ & - \\
  &$0_1 6_1 0_3$ & & $-3.788693957250$ & $0.011144489334$ & - \\
\midrule
\{1,0,0,0,1,0\}\, 0 inf &$0_1 0_5$ & $2$ & $-2.444293750583$ & $0.005887902992$ & - \\
\midrule
\{0,0,2,0,0,0\}\, 0 inf &$1_3$ & $2$ & $-1.111855930538$ & $0.001342476001$ & $0.0013{\color{red}84980817}$ \\
\midrule
\{0,1,0,1,0,0\}\, 0 inf &$0_2 0_4$ & $2$ &  $-1.560671012472$ & $0.000026982174$ & - \\
\bottomrule
\end{tabular}
\caption{All Bethe states for $L=12$ having nonzero overlap with the  N\'eel state. 
 The first column shows the string content of the Bethe states, including the number of infinite 
 rapidities. The second and third column show $2I_n$, with $I_n$ the BGT quantum numbers 
 identifying the different states, and the number $q$ of independent strings. Due to the parity invariance, 
 only positive quantum numbers are reported. In the second 
 column only the positive BGT numbers are shown. The fourth column is the Bethe state eigenenergy. 
 Finally, the last two columns show the exact overlap with the N\'eel state and the approximate 
 result obtained using the BGT equations. In the last column Bethe states containing zero-momentum 
 strings are excluded. Deviations from the exact result (digits with different colors) are 
 attributed to the string hypothesis. 
}
\label{table:Neel}
\end{table}
%%%%%%%%%%%%%%%%%%%%%%%%%%%%%%%%%%%%%%%%%%

The overlaps between all the eigenstates of the Heisenberg spin chain and the N\'eel 
state are reported in Table~\ref{table:Neel}. The first column in the Table shows 
the string content ${\mathcal S}\equiv\{s_1,\dots,s_M\}$, with $M$ being the number 
of finite rapidities. The number of infinite rapidities $N_{\infty}=L/2-M$ (see 
section~\ref{sec:1.2}) is also reported. The second column shows $2I_n$, with $I_n$ 
the Bethe-Gaudin-Takahashi quantum numbers (see section~\ref{sec:1.3}) identifying 
the $XXX$ chain eigenstates. Due to the parity invariance, only the positive quantum 
numbers are reported. The total number of independent strings, i.e., $q\equiv\sum_js_j$, 
is shown in the third column. The fourth column is the eigenstates energy eigenvalue 
$E$. The last two columns show the squared N\'eel overlaps and the corresponding result 
obtained using the Bethe-Gaudin-Takahashi equations, respectively. In the last 
column only the case with no zero-momentum strings is considered. The deviations 
from the exact diagonalization results (digits with different colors) have to be 
attributed to the string hypothesis. Notice that the overlap between the N\'eel 
state and the $S_z=0$ eigenstate in the sector with maximal total spin $S=L/2$ 
(first column in Table~\ref{table:Neel}), is given analytically as $2/B(L,L/2)$, 
with $B(x,y)$ the Newton binomial. 

\begin{figure}[t]
\begin{center}
\includegraphics[width=.9\textwidth]{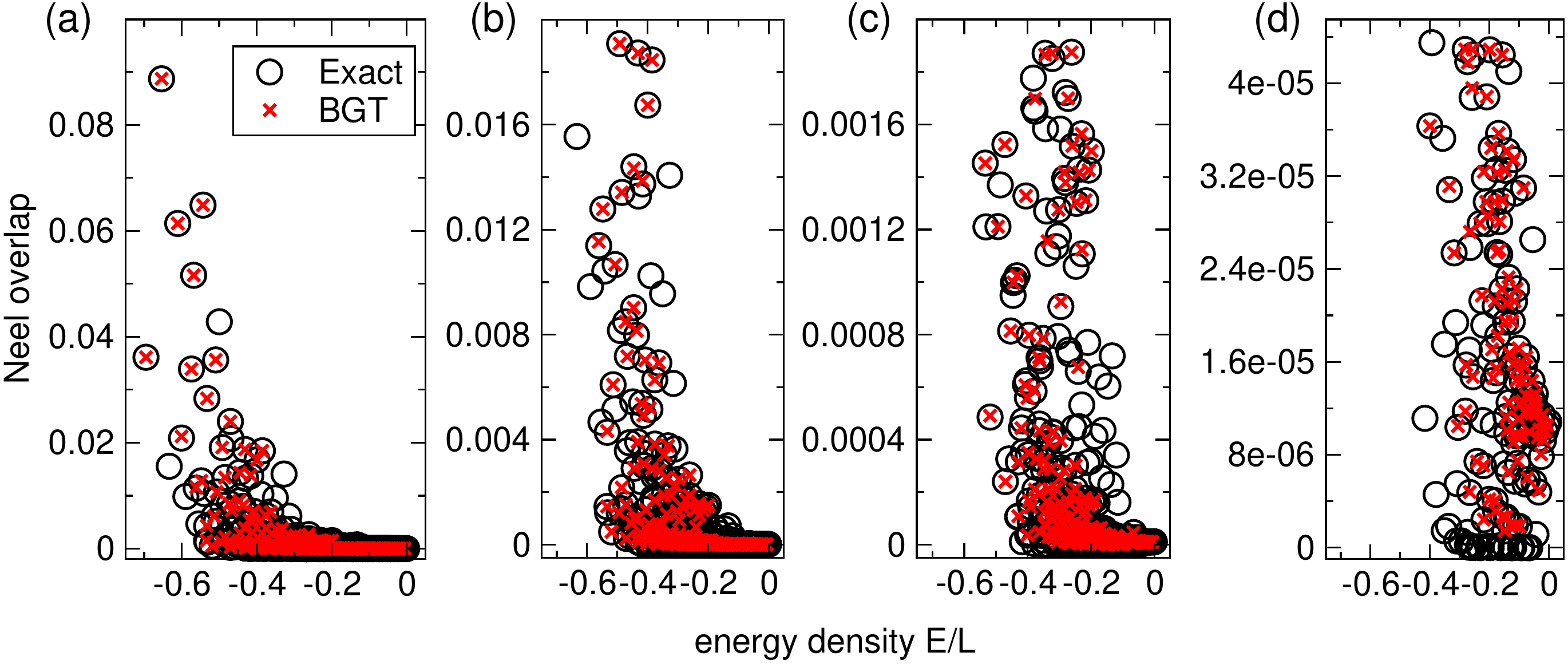}
\end{center}
\caption{ The squared overlap $|\langle N|\lambda\rangle|^2$ between the the 
 N\'eel state $|N\rangle$ and the eigenstates $|\lambda\rangle$ of the $XXX$ 
 chain with $L=20$ sites. Only non-zero overlaps are shown. In all the panels the 
 $x$-axis shows the eigenstate energy density $E/L$. The circles are the exact 
 diagonalization results for all the non-zero overlaps. The crosses are the Bethe 
 ansatz results obtained using the Bethe-Gaudin-Takahashi equations. The missing 
 crosses correspond to eigenstates containing zero-momentum strings. (a) Overview 
 of all the non-zero overlaps. (b)(c)(d) The same overlaps as in (a) zooming in 
 the regions $[0,0.2]$, $[0,0.020]$, and $[0,4\cdot 10^{-5}]$. The tiny differences 
 between the ED and the Bethe ansatz results are attributed to the string 
 deviations. 
}
\label{fig1-BGT-check}
\end{figure}

Some results for a larger chain with $L=20$ sites are reported in 
Figure~\ref{fig1-BGT-check}.  The squared overlaps $|\langle\lambda|
N\rangle|^2$ between the N\'eel state and the $XXX$ chain eigenstates $|\lambda
\rangle$ are plotted against the eigenstate energy density $E/L
\in[-\ln(2),0]$. The circles are exact diagonalization results for all 
the chain eigenstates ($382$ eigenstates), whereas the crosses denote the 
overlaps calculated using formula~\eref{Neel-ov}. Note that only the 
eigenstates with no zero-momentum strings are shown ($252$ eigenstates) 
in the Figure. Panel (a) gives an overview of all the overlaps. Panels (b)-(d) 
correspond to zooming to the smaller overlap values $|\langle N|\lambda
\rangle|\lesssim 0.02$, $|\langle N|\lambda\rangle|\lesssim 0.002$, and 
$|\langle N|\lambda\rangle|\lesssim 10^{-5}$. 
Although some deviations are present, the overall agreement between the exact 
diagonalization results and the Bethe ansatz is satisfactory, confirming 
the validity of the string hypothesis for overlap calculations.

%##################################################################
%##################################################################

%%%%%%%%%%%%%%%%%%%%%%%%%%%%%%%%%%%%%%%%%%
\begin{table}[ht]
\scriptsize
\centering
Bethe states with nonzero N\'eel overlap ($L=12$)\\[1ex]
\begin{tabular}{rrrrrr}
\toprule
String content & $2I_n$ & $q$ & E & $|\langle\lambda|MG\rangle|^2$ (exact) & $|\langle\lambda|MG\rangle|^2$ (BGT) \\[0.3em]
\toprule
\{6,0,0,0,0,0\} &$1_1 3_1 5_1$ & $6$ & $-8.387390917445$ & $0.716615769224$ & $0.716615769224$\\
\midrule
\{2,2,0,0,0,0\} &$1_1 1_2$ & $4$ & $-5.401838225870$ & $0.055624700196$ & $0.05{\color{red}4033366543}$\\  
&$3_1 1_2 $ & & $-4.613929948329$ & $0.005687428810$ & $0.005{\color{red}582983043}$\\
&$5_1 1_2 $ &  & $-3.147465758841$ & $0.002107475934$ & $0.002107{\color{red}086933}$\\
\midrule
\{3,0,1,0,0,0\} &$0_1 2_1 0_3$ & $4$ & $-6.340207488736$ & $0.205891158647$ & -\\
  &$0_1 4_1 0_3$ & & $-5.203653009936$ & $0.038832154450$ & - \\
  &$0_1 6_1 0_3$ & & $-3.788693957250$ & $0.006019410923$ & - \\
\midrule
\{1,0,0,0,1,0\} &$0_1 0_5$ & $2$ & $-2.444293750583$ & $0.000129601311$ & - \\
\midrule
\{0,0,2,0,0,0\} &$1_3$ & $2$ & $-1.111855930538$ & $0.000011727787$ & $0.00001{\color{red}2785580}$\\
\midrule
\{0,1,0,1,0,0\} &$0_2 0_4$ & $2$ &  $-1.560671012472$ & $0.000330572718$ & - \\
\bottomrule
\end{tabular}
\caption{All Bethe states for $L=12$ having nonzero overlap with the  Majumdar-Ghosh (MG) 
 state. The first column shows the string content of the Bethe states. The second and third column show 
 $2I_n$, with $I_n$ the BGT quantum numbers identifying the different states, and the number $q$ 
 of independent strings. Due to the parity invariance, we show only the positive quantum numbers. 
 In the second column only the positive BGT numbers are shown. Note that, in 
 contrast to Table~\ref{table:Neel} no states with infinite rapidities are present. The fourth column 
 is the Bethe state eigenenergy. Finally, the last two columns show the exact overlap with the MG state 
 and the approximate result obtained using the BGT equations. In the last column Bethe states containing 
 zero-momentum strings are excluded. Deviations from the exact result (digits with different colors) 
 are attributed to the string hypothesis. 
}
\label{table:mg}
\end{table}
%%%%%%%%%%%%%%%%%%%%%%%%%%%%%%%%%%%%%%%%%%

%%%%%%%%%%%%%%%%%%%%%%%%%%%%%%%%%%%%%%%%%%%%%%%%%%%%%%%%%%%%%%%%%%%%%%%%%%%
\subsection{Majumdar-Ghosh overlap}
\label{app-mg}

The overlap between all the Heisenberg chain eigenstates with the Majumdar-Ghosh state are shown in 
Table~\ref{table:mg} for the chain with $L=12$ sites. The conventions on the representation of 
the eigenstates is the same as in Table~\ref{table:Neel}. Note that in contrast with the N\'eel 
state, only the eigenstates with zero total spin $S=0$ have non zero overlap, i.e., no eigenstates 
with infinite rapidities are present, which reflect that the Majumdar-Ghosh state is invariant 
under $SU(2)$ rotations.

%%%%%%%%%%%%%%%%%%%%%% REFERENCES %%%%%%%%%%%%%%%%%%%%%%%%%%%%%%%%%%%%%%%%%%%%%%%%

\end{document}